\documentclass[aps,twocolumn,groupedaddress,longbibliography]{revtex4-1}

\usepackage{amsmath}
\usepackage[utf8]{inputenc}
\usepackage{graphicx}
\usepackage{multirow}
\usepackage{xcolor}
\usepackage{color}
\usepackage{aas_macros}
\usepackage{ulem}

\newcommand{\Dhs}{D{\cdot} \Delta h \cdot \sin^2 \theta}
\newcommand{\Dh}{D{\cdot} \Delta h}

\begin{document}

\title{Bayesian inference from gravitational waves in fast-rotating, core-collapse supernovae}

\author{Carlos Pastor-Marcos$^{1,2}$, Pablo Cerdá-Durán$^{1,3}$, Daniel Walker$^{4}$, Alejandro Torres-Forné$^{1,3}$, Ernazar Abdikamalov$^{5,6}$, Sherwood Richers$^{7}$, and Jos\'e A.~Font$^{1,3}$}
\affiliation{$^1$ Departamento de Astronom\'ia y Astrof\'isica, Universitat de Val\`encia, Dr. Moliner 50, 46100, Burjassot (Valencia), Spain}
\affiliation{$^2$ Institut für Theoretische Physik, Ruprecht-Karls-Universität Heidelberg, Philosophenweg 16, 69120 Heidelberg, Germany}
\affiliation{$^3$ Observatori Astron\`omic, Universitat de Val\`encia, Catedr\'atico Jos\'e Beltr\'an 2, 46980, Paterna, Spain}
\affiliation{$^4$ Department of Physics, Blackett Laboratory, Imperial College, London SW7 2AZ, UK}
\affiliation{$^5$ Department of Physics, School of Sciences and Humanities, Nazarbayev University, Astana
010000, Kazakhstan}
\affiliation{$^6$ Energetic Cosmos Laboratory, Nazarbayev University, Astana 010000, Kazakhstan}
\affiliation{$^7$ Department of Physics and Astronomy, University of Tennessee, Knoxville, TN 37996-1200, USA}

\date{\today}

\begin{abstract}
Core-collapse supernovae (CCSNe) are prime candidates for gravitational-wave detectors. The analysis of their complex waveforms can potentially provide information on the physical processes operating during the collapse of the iron cores of massive stars. In this work we analyze the early-bounce rapidly rotating CCSN signals reported in the waveform catalog of Richers et al 2017. This catalog comprises over $1800$ 
axisymmetric simulations extending up to about $10$~ms of post-bounce evolution. It was previously established that for a large range of progenitors, the amplitude of the bounce signal, $\Dh$, is proportional to the ratio of rotational-kinetic energy to potential energy, $T/|W|$, and the peak frequency, $f_{\rm peak}$, is proportional to the square root of the central rest-mass density, $\sqrt{\rho_{\rm c}}$. In this work, we exploit these relations to suggest that it could be possible to use such waveforms to infer protoneutron star properties from a future gravitational wave observation, but only if the distance and inclination are well known and the rotation rate is sufficiently low. Our approach relies on the ability to describe a subset of the waveforms in the early post-bounce phase in a simple form -- a master waveform template -- depending only on two parameters, $\Dh$ and $f_{\rm peak}$. We use this template to perform a Bayesian inference analysis of waveform injections in Gaussian colored noise for a network of three gravitational wave detectors formed by Advanced LIGO and Advanced Virgo. 
We show that, for a galactic event ($D\sim10$~kpc), it is possible to recover the peak frequency and amplitude with an accuracy better than $10\%$ for $\sim 80\%$ and $\sim 60\%$ of the signals, respectively, given known distance and inclination angle.
However, inference on waveforms from outside the Richers catalog is not reliable, indicating a need for carefully verified waveforms of the first $10\,\mathrm{ms}$ after bounce of rapidly rotating supernovae of different progenitors with agreement between different codes.

\end{abstract}
\maketitle


\section{Introduction}
\label{sec:intro}

Core-collapse supernova (CCSN) explosions are prime astrophysical sources of transient gravitational waves (GWs; see e.g.~\cite{Kotake2017,Abdikamalov:2022} and references therein). If the explosions occur sufficiently close (i.e.~inside the galaxy or within the local system of satellite galaxies) GWs from CCSNe could be within reach \cite{observing,Abbott:2020,Szczepanczyk2021,Marek:2023,Mezzacappa2023,Lagos2023} of the current network of ground-based observatories Advanced LIGO \cite{aLIGO}, Advanced Virgo \cite{aVirgo}, and KAGRA \cite{KAGRA}.
The most recent optically-targeted search for GW transients associated with CCSNe within a source distance of 30 Mpc using data from the third observing run of Advanced LIGO and Advanced Virgo has been reported in~\cite{Marek:2023}. No significant GW candidate was announced. The rate of galactic CCSNe is $1$-$3$ per century (see \cite{Gossan:2016,Abbott:2020} and references therein) and about $1$-$10\%$ of all events may have significant rotation (see discussion in \cite{Raynaud2022}). The mechanism by which rapidly rotating progenitor cores are formed is also poorly understood, since magnetic and turbulent stresses can drive the star toward slower rigid rotation (e.g., \cite{Heger2005,varma20233d}). Despite their low intrinsic event rate, CCSNe hold great scientific interest as the analysis of their complex waveforms can potentially provide valuable information about the underlying physical processes operating during the gravitational collapse of the iron cores of massive stars. 

At the onset of collapse the iron core, where no nuclear burning is taking place, is very close to axisymmetry. If the core is rotating sufficiently quickly, the dominant deviation from spherical symmetry below the silicon burning shell is produced by its rotation, which imprints an oblate shape into the core, contributing to its mass quadrupole. The collapse produces an acceleration of the infalling material and, therefore, a non-zero second time derivative of the mass quadrupole, inducing GW emission. The peak GW amplitude is reached right before bounce, at which time the nuclear equation of state (EOS) stiffens. This sharply halts the collapse and produces a small re-expansion and a series of oscillations of the newly formed proto-neutron star (PNS), which are visible in the GW signal for a few milliseconds. The maximum GW strain is related to the degree of oblateness of the core at bounce, which is in turn related to the rotation rate; hence, for non-rotating progenitors the GW emission at bounce is zero. Numerical simulations have shown that the frequency of the bounce oscillations is in the range $100$-$1000$~Hz \cite{Dimmelmeier:2008,richers2017equation}. These oscillations are the result of the non-linear excitation of an axisymmetric $f$-mode of the PNS whose frequency depends on the local sound speed in this region \cite{Fuller2015}. Beyond $\sim10\,\mathrm{ms}$ after bounce, the growth of hydrodynamical instabilities breaks axisymmetry (see e.g.~\cite{Janka2017}) and the GW signal, with a typical duration of $\sim 1$~s, becomes stochastic. In this work we focus our attention in the bounce GW signal. 

Since the seminal numerical study of~\cite{Zwerger:1997}
the bounce signal has been extensively studied in a number of works~\cite{Dimmelmeier:2002,Dimmelmeier:2008,Ott:2012,Abdikamalov:2014,Fuller2015,richers2017equation,Edwards2021}. Richers et al. \cite{richers2017equation} performed over $1800$ 2D (axisymmetric) numerical simulations of CCSNe that include the collapse phase and up to about $10$~ms of the post-bounce evolution. For such short post-bounce timescales axisymmetry is a valid approximation because all possible non-axisymmetric instabilities (such as convection and the standing accretion shock instability) develop in longer timescales. The limiting factor is probably the occurrence of prompt convection that for the typical values of the Brunt-Väisälä frequency ($\sim 100$~ms) may develop in timescales of $\gtrsim 10$~ms. Furthermore, the post-bounce timescale covered in~\cite{richers2017equation} is much shorter than the characteristic timescale in which the PNS cools due to diffusing neutrinos ($\gtrsim 100$~ms), which renders unnecessary the incorporation of neutrino transport for an accurate modelling, although the equation of state is still sensitive to deleptonization of infalling matter by neutrino emission. Regarding GW emission, the axisymmetry of the system implies that the GW signal is completely polarized. The study of~\cite{richers2017equation} has shown that, for a $12 M_\odot$ progenitor and a large range of EOS, the amplitude of the GW signal at bounce is proportional to the ratio of rotational-kinetic energy to gravitational potential energy, $T/|W|$, and the peak frequency is proportional to the square root of the central rest mass density, $\sqrt{\rho_{\rm c}}$. These relations break for sufficiently large rotation rates, $T/W>0.06$, in which case the centrifugal forces have an impact in the dynamics of the collapse and the bounce. These high rotation rates are typically predicted only for progenitors of long gamma-ray bursts in low-metallicity enviroments (see e.g.~\cite{Woosley2006}). The relative rarity of extremely energetic supernovae (e.g., \cite{Guetta_2007}) suggest that much more common are progenitors in which only a part of the rotation is retained by the core leading to values of $T/|W|$ well below $0.06$ (see e.g.~\cite{Heger2005}). Therefore, for the most common rotating CCSNe these two phenomenological laws could allow us to infer some PNS properties from a future GW observation, which is the main aim of this investigation.

Efforts to infer the rotational properties of newborn PNS from the GW signal have been attempted in previous works. In~\cite{Hayama:2008} the authors tried to estimate the distribution of angular momentum in the PNS from the sign of the second peak in the bounce signal. A matched-filtering analysis method to infer the total angular momentum of the core with $20-30\%$ accuracy using CCSN bounce signals injected in Gaussian detector noise was developed by~\cite{Abdikamalov:2014}. The main caveats of this work are the choice of source orientation (optimal), the use of simulation waveforms as templates (instead of parametric templates) and the inability of the method to estimate errors associated with the measurements. Additionally, several works \cite{Rover2009,Logue2012,Powell2016,Powell2017} have developed data analysis pipelines based on principal component analysis (PCA) of the signal, capable of performing Bayesian model selection. In particular, for sufficiently close sources, these investigations were able to distinguish between neutrino-driven and magnetorrotational explosions, allowing to assess the presence of rotation in CCSNe. Similar classification approaches and outcomes, but based on machine-learning techniques, have been developed by \cite{Chan2020,Ainara:2022}. For a limited set of simulations, \cite{Pajkos2021} were able to infer details on the progenitor core structure combining information from the bounce and post-bounce GW signal. Ref. \cite{Afle2021} performed a PCA decomposition of the simulation catalog of \cite{richers2017equation} to determine the dominant features of the waveforms and create a map between the measured properties of the waveforms and the physical properties of the progenitor star (i.e.~$T/|W|$ and peak frequency). Their analysis shows that this is possible for galactic CCSNe with current GW detectors and up to $50$~kpc with 3rd generation detectors. As in the case of \cite{Abdikamalov:2014}, they use optimally oriented sources for their analysis, which, as we show in this work, limits the ability of the results to generalize to observations with unknown orientation. Finally, an alternative signature of rapid rotation is the presence of circular polarization in the post-bounce signal~\cite{Hayama2016}, which could be detectable within $5$~kpc by current detectors~\cite{Chan2021}.

The starting assumption of our work is the multi-messenger detection of a nearby CCSN. The first indication from the occurrence of one such event would be provided by the simultaneous 
observation of neutrinos and GWs. Neutrino detectors such as SuperK~\cite{SuperK} and IceCube~\cite{IceCube} are able to detect MeV neutrinos from CCSNe at distances of about $100$~kpc. In the case of an event, the network of neutrino detectors would emit an alert by means
of the Supernovae Early Warning System 
(SNEWS~\cite{SNEWS,SNEWS2}). These alerts provide estimations of the time of bounce in $10$~s windows. However, a detailed analysis of the data from neutrino detectors should allow to estimate the time of bounce within $10$~ms~\cite{Pagliaroli:2009,Halzen:2009}. On the other hand, online GW burst searches such as Coherent Wave Burst (CWB~\cite{Klimenko2016}) used in the current LIGO-Virgo-KAGRA network of advanced GW detectors are capable of detecting GWs from nearby CCSNe for the case of neutrino-driven explosions and up to $\sim100$ ~kpc for extremely fast-rotating progenitors~\cite{Szczepanczyk2021}. In the case of a successful GW detection, it should be possible to obtain an accurate measurement of the time of bounce within ms. Later on, on a timescale of minutes to days after bounce, the electromagnetic signature of the supernovae would allow an accurate determination of the sky location and of the distance to the source if linked to a known presupernova star. Neutrino detections can provide complementary direction and distance estimates (e.g.~\cite{SNEWS2,Kachelriess2005Exploiting}).

For the purpose of this paper, we will assume that a GW detection of a CCSN has been made and that an accurate enough measurement of the time of bounce, sky location and distance was possible. Under these circumstances, our goal is to develop a data analysis framework to infer the properties of the characteristic bounce signal, signature of the presence of rotation, as a first step towards inferring the rotation rate and mean density of the PNS. For this purpose, in Section~\ref{sec:MT} we develop a parametric waveform template based on the numerical CCSN simulations of~\cite{richers2017equation}. In section~\ref{sec:CCSNWaveforms} we describe the additional numerical waveforms used for testing.
In Section~\ref{sec:method} we introduce the Bayesian method that we use to perform the parameter estimation of the signal. The results of this analysis for the case of a three-detector network with Gaussian colored noise are shown in Section~\ref{sec:results}. Finally, our conclusions are presented in Section~\ref{sec:conclusions}.


\begin{figure}[t]
 \includegraphics[width=0.49\textwidth]{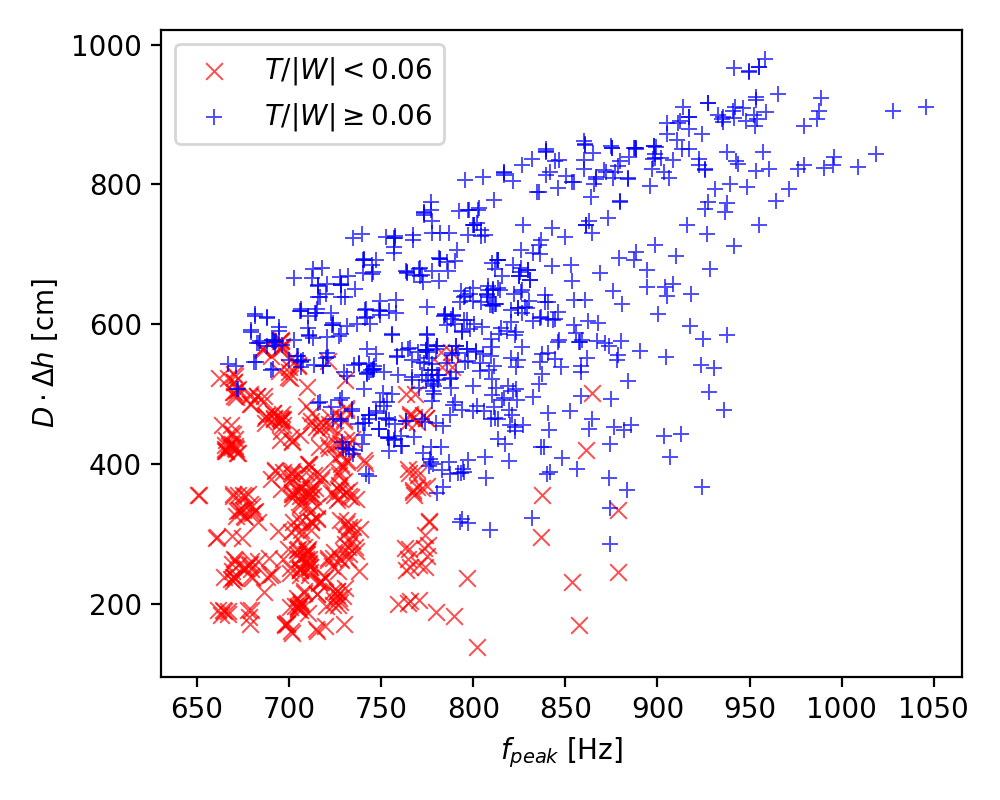}
 \caption{Waveform amplitude, $\Dh$, vs peak frequency, $f_{\rm peak}$, for all models in the waveform catalog of \cite{richers2017equation}. Red (blue) markers indicate models included (excluded) in our analysis. The two kind of models occupy regions of the parameter space that are mostly disjoint.} 
 \label{fig:RichersRegimesToverW}
\end{figure}

\begin{figure*}[t]
 \includegraphics[width=0.49\textwidth]{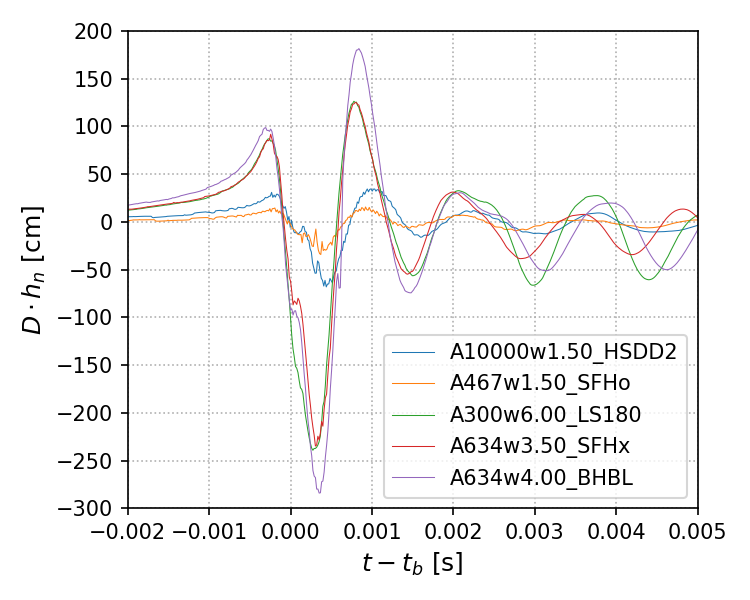}
 \includegraphics[width=0.49\textwidth]{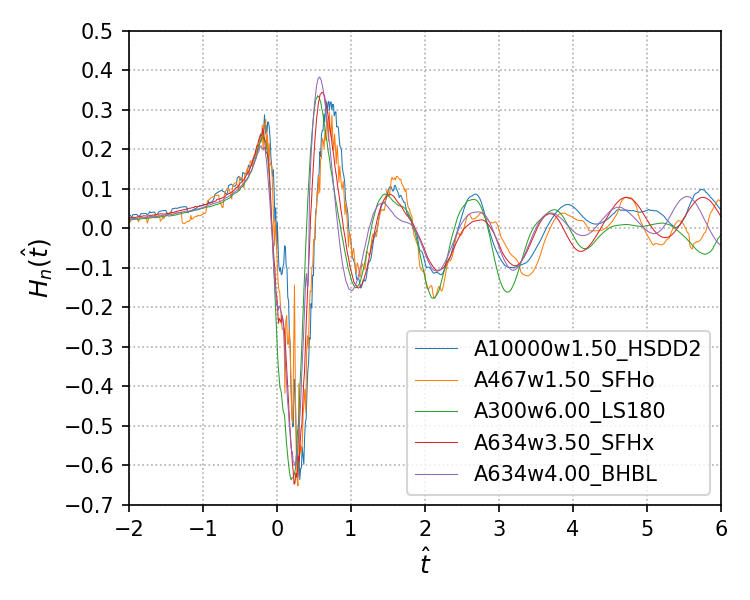}
 \caption{Strain as a function of time (left panel) and normalized strain as a function of the normalized time (right panel) for five ilustrative examples. These waveforms have been selected from simulations that are as different as possible from each other to display the diversity of waveforms.
 The right panel shows the universality of the shape of the normalized waveform. The simulation model names have the form {\texttt AXwY\_Z}, indicating \texttt{X} the degree of differential rotation, \texttt{Y} the value of $\Omega_0$ in rad/s and \texttt{Z} the EOS used (see more details in \cite{richers2017equation}).}
 \label{fig:normalization}
\end{figure*}

The work of \cite{richers2017equation} shows that for a large range of rotation rates ($T/W<0.06$), the two most relevant parameters characterizing the waveform at bounce are the peak amplitude (normalized by the distance to the source), $\Dh$, and the peak frequency, $f_{\rm peak}$. The peak amplitude is measured as the difference between the highest and lowest points in the bounce signal strain assuming optimal orientation. Correspondingly, the peak frequency is the largest frequency measured from the signal in the first $6$~ms following the bounce. We show in this section that it is possible to describe the waveforms in the early post-bounce phase in a simple form -- a master waveform template -- depending only on these two parameters: $\Dh$ and $f_{\rm peak}$.

\section{Bounce signal characterization}
\label{sec:MT}

\subsection{Waveform selection and renormalisation}
\label{sec:renorm}

We employ data from the 1824 numerical-relativity simulations carried out in~\cite{richers2017equation}. (Note that this catalog is publicly available in \cite{richers_sherwood_2016_201145}.) These simulations were performed using the CoCoNuT code \cite{Dimmelmeier:2002} in general relativity (XCFC approximation \cite{Cordero-Carrion2009}) and a leakage scheme \cite{Ott:2012} for the neutrino transport. All simulations were performed with the same $12 M_\odot$ progenitor star, but different initial rotational profile and rate, and using $18$ different equations of state (EOS) (see \cite{richers2017equation} for additional details). The simulations focused on the bounce signal, therefore they were run only $50$~ms after bounce.

For each simulation, the catalog provides the strain in units of distance for an optimally oriented source, $D \cdot h_+^{\rm opt}$ as a function of time after core bounce $t-t_{\rm b}$, the peak frequency of the post-bounce GW oscillations $f_{\mathrm{peak}}$, the ratio of rotational kinetic energy to gravitational potential energy of the inner core at bounce $T/|W|$, and the maximum initial pre-collapse rotation rate $\Omega_0$. The minimum and maximum values of the first and second GW strain peak  allow us to obtain the peak amplitude, $\Dh$. Since the system is axisymmetric, the strain for different source orientations can be computed simply as
\begin{eqnarray}
 h_+ &=& h_+^{\rm opt} \,\sin^2\theta\,, \nonumber \\
 h_\times &=& 0\,,
\end{eqnarray}
where $\theta$ is the inclination angle between the rotational axis of the core and the observer's line of sight. Note that in the coordinate system of the source there is no dependence on the azimuthal angle because of axisymmetry. Note also that the strain in the coordinate system of the observer depends on an additional polarization angle $\psi$ measuring the orientation of the projection of the rotation axis on the sky with respect to the celestial meridian.

Some of the 1824 simulations performed 
in~\cite{richers2017equation} do not undergo core collapse within the simulation time as a consequence of the initial values of the parameters describing the progenitors. For those models, the rotational kinetic energy is large enough to yield sufficient centrifugal support at the onset of collapse to prevent them from collapsing. Those waveforms are listed in Table III of~\cite{richers2017equation} and they have been excluded from our analysis. In addition, \cite{richers2017equation} observed that the waveforms of extremely fast-rotating cores show no specific trend with $\Dh$ and $f_{\mathrm{peak}}$. However, if rotation is sufficiently slow ($T/W<0.06$), $\Dh$ increases linearly with $T/|W|$ (see Fig.~6 of~\cite{richers2017equation}), due to the quadrupolar deformation of the core induced by rotation. Taking this observation into account we restrict the number of waveforms of our analysis to those for which $0.00<T/|W|<0.06$. We note that within this limit there is no particular dependence of $f_{\mathrm{peak}}$ with $T/|W|$. As discussed in Section~\ref{sec:intro}, most CCSN progenitors are expected to be in this range, and only those capable of producing long gamma-ray bursts may be above the upper limit. However, even with this limitation it could be possible to devise algorithms able to determine whether a particular observation is in this linear regime or not, based purely on observations. Fig.~\ref{fig:RichersRegimesToverW} shows the values of $f_{\mathrm{peak}}$ and $\Dh$ for all the models in \cite{richers2017equation}. Models with $T/|W|<0.06$ (used in our analysis) occupy a region of the parameter space mostly disjoint to models with $T/|W|>0.06$ (excluded of our analysis). Therefore, measuring $f_{\mathrm{peak}}$ and $\Dh$ could help determining whether a signal is in the linear regime or not. We refer to this selection of waveforms as Richers et al. catalog, hereafter.

\begin{figure*}[t]
 \includegraphics[width=0.48\textwidth]{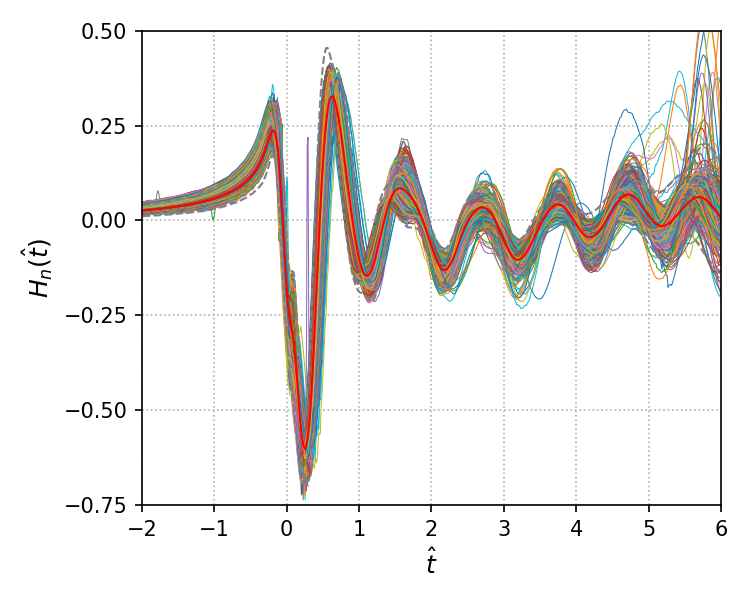}
 \includegraphics[width=0.48\textwidth]{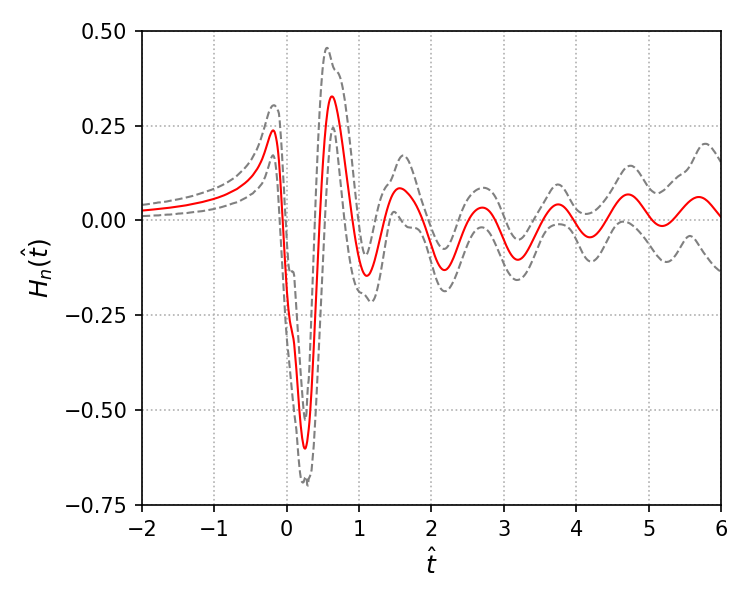}
 \caption{Master template. Left panel shows $H_{\rm T}(\hat t)$ for the master template (red line) together with the 402 waveforms used for its construction. Right panel shows the same master template with the $2$-$\sigma$ ($\sim 95\%$) confidence interval computed using $\Sigma (\hat t)$ (dashed lines). } 
 \label{fig:master-template}
\end{figure*}

Each of the Richers et al. catalog waveforms has been extracted from a particular simulation. The waveform morphology at bounce, while similar, is still somewhat diverse (see left panel of Fig.~\ref{fig:normalization} for a few examples). The work of~\cite{richers2017equation} suggests that the main 
parameters describing the waveforms are $\Dh$ and $f_{\rm peak}$. Hence, it seems natural to normalize all the waveforms by these two parameters to try to suppress as much as possible all implicit dependencies for each individual waveform. To do this, we first align the time of bounce for all waveforms and renormalize the value of the strain with $\Dh$. Due to the linear dependence of $\Dh$ on $T/|W|$ the result of this rescaling is the effective elimination of the dependence of each waveform on rotation. 
Additionally, we rescale the time by multipliying by $f_{\rm peak}$. Given the approximate linear dependence of $f_{\mathrm{peak}}$ with $1/\sqrt{\rho_{\rm c}}$ ($\rho_{\rm c}$ being the mean central density within the $0.2$~ms after bounce), this rescaling effectively removes the dependence of the waveform on $\rho_{\rm c}$.

Let us denote by $h_n(t)$ the strain of the $n$-th waveform in the catalog at optimal orientation and at a distance of $10$~kpc, with $n=1,...,N_T$ and $N_T$ the total number of waveforms. Its normalized version is thus defined as:
\begin{equation}
 H_n(\hat{t})=\frac{D{\cdot}h_n( (t - t_{\rm b}) \cdot f^n_{\rm peak})}{\Dh_n},\qquad \hat{t}\equiv (t - t_{\rm b})\cdot f_{{\rm peak}}^n,
\end{equation}
where $\Dh_n$ and $f_{\rm peak}^n$ denote the corresponding values for the $n$-th waveform. 
Note that we keep the $D$ explicitly in front of the strain quantities to have combinations of quantities that are source properties independent of the distance. The right panel of Fig.~\ref{fig:normalization} shows the normalized waveforms, $H_n(\hat{t})$, resulting from rescaling the waveforms in the left panel. Once the dominant dependencies of the waveform are eliminated, the shape of the main peak and the first few oscillations match closely. At late times, differences become larger due to the onset of hydrodynamical instabilities in the PNS (e.g.~convection) that have a stochastic behaviour. Nevertheless, overall the overlap is acceptable for such a simple two-parameter normalization.

The most significant distinction between different normalized waveforms is the presence of spurious, high-frequency
oscillations in some of them, in particular for models with slow rotation rates. The source of these oscillations is numerical noise in the calculation of the GW signal using the quadrupole formula. In the limit of decreasing rotation, $\Omega_0 \to 0$, the object becomes spherically symmetric so the strain should vanish. However, computationally this implies perfect numerical cancellation of the integrals over the source necessary to evaluate the quadrupole formula, which inevitably leads to numerical noise. This noise, which is always present, is more evident in signals with lower amplitudes because the strain is comparable to the noise. The normalization factor $1/(\Dh)$ amplifies the numerical noise for the case of slow rotation progenitors. By visual inspection we have checked that this numerical noise ceases to be relevant for cases with $\Omega_0 \geq 3$~rad/s. Therefore, all waveforms not fulfilling this requirement have also been excluded from our analysis.

\subsection{Master Template}
\label{sec:MasterTemplate}

After the model selection discussed in the previous section we are left with a total of $N_T=402$ different waveforms whose normalized strain, $H_n(\hat t)$, depends weakly on the particular simulation. We use the average value of all $H_n (\hat t)$ to create a master function $H_{\rm T}(\hat t)$ that can be utilized for constructing waveform templates through the straightforward procedure of rescaling by $\Dh$ and $f_{\rm peak}$ (in addition to the dependence on $D$ and $\theta$).

The first necessary step is thus the averaging of all waveforms. The time coordinate of the signals in the waveform catalog is different depending on the simulation and after the rescaling this becomes more evident in the normalized signals. Therefore, before doing any averaging it is necessary to perform an interpolation to a common grid
of normalized times, $\hat t$. We create this common grid within the interval $\hat t \in [-4,6]$, since for $\hat t<-4$ there are no events of interest and for $\hat t>6$ the divergence of the waveforms is too severe. We use a cubic interpolation with 1000 points in the interval.

For all waveforms we compute the average and standard deviation of $H_n$ for each value of $\hat t$, obtaining a triplet $\hat{t}-\bar{H}-\Sigma$ that defines the master waveform:
\begin{align}
 &H_T(\hat{t}) \equiv \frac{1}{N_T}\sum_{n=1}^{N_T}H_n(\hat{t})\,,\\
 &\Sigma^2(\hat{t}) \equiv \frac{1}{N_T}\sum_{n=1}^{N_T}|H_n-H_T|^2\,.
\end{align}
The standard deviation $\Sigma$ accounts for the error made when performing inference over the master template. However, all our estimations have been done without considering this error. 
The master waveform template is shown in Fig.~\ref{fig:master-template} together with the $2\Sigma$ interval corresponding to a $\sim 95\%$ confidence level.

The master template is expected to apply to a wide range of parameters and to allow us to construct a parameterized template, suitable for performing Bayesian inference. Our phenomenological waveforms depend on $\Dh$, $f_{\mathrm{peak}}$, $D$ and $\theta$, and can be expressed as:
\begin{eqnarray}
 h_+ (t) = H_{\rm T}(t \cdot f_{\rm peak}) \, \frac{\Dhs}{D}, \hspace{0.5cm}
 h_\times (t) = 0. \label{eq:template}
\end{eqnarray}
We note that to keep explicit the dependence on the combination $\Dh$ we have to divide by $D$. Additionally, for a fixed distance and $f_{\rm peak}$, the strain amplitude is proportional to the combination $\Dhs$. In the next section we discuss the importance of this combination as a parameter.

\section{Core collapse waveforms}
\label{sec:CCSNWaveforms}

\begin{table*}[]
 \centering 
 \caption{List of CCSN simulations used for testing the Bayesian inference algorithm. $M_{\rm ZAMS}$ refers to the zero-age main-sequence mass of the progenitor star. }
 \label{tab:CCSNModels}
 \begin{tabular}{llccccccc}
  \hline\hline
   Source & $\qquad$ & Model & Dim.& EOS & $M_{\rm ZAMS}$ [$M_\odot$] & $\Omega_0$ [rad/s] & $\Dh$ [cm] & $f_{\rm peak}$[Hz] \\ 
   \hline 
   Morozova et al. (2018) &\cite{Morozova2018} & M13\_SFHo\_rotating & 2D & SFHo & $13.0$ & $0.2$ & $3.81$ & $538.0$ \\
   \hline
   Bugli et al. (2020) & \cite{Bugli2020}
   & l1\_r2M & 2D& LS220 & 35.0 & 2.0 & 49.9 & 821.3\\
   & & l2\_r2M & 2D& LS220 & 35.0 & 2.0 & 48.3 & 800.1 \\
   & & l3\_r2M & 2D & LS220 & 35.0 & 2.0 & 46.7 & 803.1 \\
   & & l4\_r2M & 2D & LS220 & 35.0 & 2.0 & 45.7 & 802.2 \\
   & & hydro\_2d &  2D& LS220 & 35.0 &  2.0 & 42.5 & 819.0 \\
   & & hydro\_3d &  3D & LS220 & 35.0 &  2.0 & 163.8 & 894.6 \\
   \hline
   Obergaulinger \& Aloy (2020) &\cite{Obergaulinger2020} 
   & 35OC-RO  & 2D& LS220  & 35.0 & 2.0 & 71.0 & 734.9 \\
   & & 35OC-RO2 & 2D& LS220  & 35.0 & 2.0 & 42.6 & 816.5 \\
   & & 35OC-Rp2 & 2D& LS220 & 35.0 & 2.0 & 42.6& 640.7 \\
   & & 35OC-Rp3 & 2D& LS220 &  35.0 & 2.0& 40.1 &713.6 \\
   & & 35OC-Rp4 & 2D & LS220 &  35.0 & 2.0 & 44.5 & 709.7 \\
   & & 35OC-Rs & 2D & LS220 & $35.0$ & $2.0$ & $54.3$ & $797.4$ \\
   & & 35OC-RRw &  2D & LS220 & 35.0 & 4.0 & 95.0& 717.6 \\
   \hline
   Powell et al. (2020) & \cite{Powell2020} 
    &m39 &  3D & LS220 & 39.0 &  0.54 & 142.0 & 739.1\\
   \hline
   Obergaulinger \& Aloy (2021) & \cite{Obergaulinger2021}
   & O & 3D & LS220 & 35.0 & 2.0& 200.4 & 809.1 \\
   & & W & 3D & LS220 &  35.0 & 2.0 & 237.1 & 734.6 \\
   \hline
   Pan et al. (2021) & \cite{Pan2021} 
   & s40\_FR & 3D & LS220 & 40.0 & 1.0 & 212.6 & 897.5 \\
   \hline 
   Obergaulinger et al. (2022) & \cite{Obergaulinger2022}
   & A08 & 2D & SFHo & 8.0 & 0.2 & 2.7 & 966.9 \\
   & & A13  & 2D & SFHo & 13.0 & 0.2 & 4.1 & 655.3 \\
   & & A20 & 2D & SFHo & 20.0 & 0.2 & 5.0 & 816.7 \\
   & & A39  & 2D & SFHo & 39.0 & 0.2 & 7.3 & 964.3\\
   \hline\hline
 \end{tabular}
\end{table*}

In addition to the waveforms from \cite{richers2017equation} we use in this work a series of GW templates from other authors that serve as a test base for our Bayesian inference procedure. The waveforms considered correspond to some models from \cite{Morozova2018,Bugli2020,Obergaulinger2020,Powell2020,Obergaulinger2021,Pan2021,Obergaulinger2022} and their details are reported in Table~\ref{tab:CCSNModels}. This selection is not meant to be complete among all possible existing numerical simulations of fast rotating cores and serves only as a testbed of our approach. However, note that the minimum value of $\Omega_0$ that we admit from the Richers catalog is $\Omega_0=3\,\mathrm{rad/s}$, but the maximum value for the CC waveforms is $\Omega_0=2\,\mathrm{rad/s}$. Although none of these supernovae are rotating extremely quickly and we expect the template to be suitable throughout, future multidimensional simulations with more rapid rotation will benefit this analysis.

The simulations from \cite{Morozova2018} were performed using the FORNAX code \cite{Skinner2019} with a mutigroup M1 neutrino transport. \cite{Bugli2020,Obergaulinger2020,Obergaulinger2021,Obergaulinger2022} used the AENUS-ALCAR code \cite{Just2015} with a mutigroup M1 neutrino transport. \cite{Pan2021} uses the FLASH code with IDSA neutrino transport \cite{Fryxell2000,Dubey2008}. \cite{Powell2020} uses the CoCoNuT code \cite{Dimmelmeier:2002} and the fast multigroup neutrino transport method \cite{Muller2015}. All the codes, except CoCoNuT, use pseudo-relativistic gravity~\cite{Marek2006}.
Gravity in CoCoNuT is general relativistic using the XCFC approximation \cite{Cordero-Carrion2009}. Gravitational waves were extracted using the quadrupole-formula in all cases \cite{Finn_1990,Blanchet_1990}.

Among the models there are full three-dimensional simulations (3D) and axisymmetric simulations (2D).
Different from the simulations from \cite{richers2017equation}, these models include the post-bounce evolution for several hundred milliseconds, showing additional features in the gravitational wave signal such as g-mode excitation and, in some cases, the standing accretion shock instability. However, these occur on timescales longer than the window used in the analysis of the bounce signal and have little consequence for the present study. The values of $f_{\rm peak}$ and $\Dh$ are estimated from the waveforms following the same procedure outlined in \cite{richers2017equation}.


\section{Methodology}
\label{sec:method}

\subsection{Bayesian inference}
\label{sec:bayes}

The inference of parameters of our CCSN waveforms is achieved through Bayesian analysis, which is briefly reviewed next.

Let $H$ be an uncertain hypothesis (the presence of a bounce signal that can be modelled by our master template), $\Theta$ the set of source parameters that we want to constrain or estimate and $d$ the observational data. Bayes' theorem states that (see e.g.~\cite{Veitch_2015,Christensen2022})
\begin{equation}
p(\Theta|d,H)=\frac{p(\Theta|H)\,p(d|\Theta,H)}{p(d|H)}, 
\label{eq:BayesTheoremGW}
\end{equation}
where $p(\Theta|d,H)$ is the probability distribution of the parameters $\Theta$ given both the data and the hypothesis (posterior distribution), $p(\Theta|H)$ gives the expectation of the parameters given the hypothesis (prior distribution), $p(d|H)$ is the expectation of the observed data given the hypothesis (evidence), and $p(d|\Theta,H)$ is the probability distribution of observing the data, given the parameters and hypothesis 
(the likelihood, $\mathcal L$).

We use the Bayesian inference library \texttt{Bilby} \cite{Ashton_2019}, which is an open-source, MIT-licensed \texttt{Python} code developed by the LVK Collaboration, that provides the parameter estimation infrastructure that we need to analyze our data. We model the observed signal at each detector, $s(t)$, as a linear combination of the actual GW signal, $h(t)$, and the detector's noise, $n(t)$; i.e.~$s(t) = h(t) + n(t)$. Based on this assumption, we use a Gaussian likelihood of the form \cite{Christensen2022}
\begin{equation}
  \log \mathcal{L} = -\frac{2}{T} \sum_{\rm IFOs}\sum_{i=1}^{N/2} \frac{|\hat s_i - \hat h_i|^2}{S_{n\,i}},
  \label{eq:likelihood}
\end{equation}
$T$ being the duration of the signal, $N$ the number of samples, $\hat s_i$ the Fourier transform of the signal, $\hat h_i$ the Fourier transform of the template and $S_{ni}$ the power spectral density (PSD) of the detector's noise. The summation sign {\it IFOs} means a sum over all GW interferometers in the network. The Fourier transforms are computed from the discrete Fourier transform as
\begin{equation}
\hat{h}_k = \sum_{j=1}^{N} h(t_j) e^{-\frac{2\pi i}{N} (j-1) (k-1)} \Delta t \quad ; \quad k = 1, ..., N.
\end{equation}

We consider the waveform template model described in Section \ref{sec:MT} (see 
Eq.~\eqref{eq:template}), which, in principle, depends on four intrinsic parameters: $\{ \Dh, f_{\rm peak}, D, \theta \}$.
In a general situation, the actual GW strain observed in a detector depends on the source location and on the polarization of the waves, which is taken into account by defining the so-called antenna response, $F(\alpha,\delta,\psi)$, such that
\begin{equation}
h = F_+(\alpha,\delta,\psi)h_+ + F_{\times}(\alpha,\delta,\psi)h_{\times}.
\end{equation}
Here, $\alpha$ and $\delta$ are the right ascension and declination of the source, respectively, and $\psi$ is the polarization angle, which describes the orientation of the projection of the progenitor's rotation axis onto the plane of the sky (see discussion in Section~\ref{sec:renorm}). \texttt{Bilby} combines the source location and orientation with detector characteristics to calculate $F_+$ and $F_\times$, and therefore determine each detector's observed waveform.

For our analysis we will consider that the sky location ($\alpha$ and $\delta$), the distance $D$ and the time of bounce $t_{\rm b}$ are known to arbitrary accuracy (see discusion in Section~\ref{sec:intro}). Hence, they are kept fixed to the values of the injection. The combination $\Dhs$ appears in Eq.~\eqref{eq:template} as proportional to the waveform amplitude, so the parameters $\Dh$ and $\theta$ will be degenerate. Therefore, we use the combination $\Dhs$
for our parameter estimation. We discuss in Section~\ref{sec:conclusions} possible ways of breaking this degeneracy. This means that we will perform our Bayesian inference for three unknown parameters $\Theta = \{\Dhs, f_{\rm peak}, \psi \}$. In practice this is done by performing inference for $\Dh$ at fixed $\theta$ (different for each injection) and then computing $\Dhs$ itself.
For $\Dh$ we consider a uniform prior for its logarithm in the interval $[0.01,1000]$~cm (recall that $\Dh$ is intrinsic to the source and does not depend on distance). This range spans from a value sufficiently small to represent the non-rotation limit to the upper limit found in the simulations by \cite{richers2017equation}. For $f_{\rm peak}$ we use a uniform prior in the range $[300, 1500]$ Hz, that includes the range found in the simulations by \cite{richers2017equation}, $\sim[650,1050]$~Hz, with a buffer of several $100$~Hz to account uncertainties about the EOS. For $\psi$ we use a uniform prior in the range $[0,\pi]$.

All waveforms of our catalog are injected in simulated, zero-mean, coloured Gaussian noise using the power spectral densities (PSD) of Advanced LIGO and Advanced Virgo~\cite{Abbott2018, SensitivityCurve}. We use the three detector network formed by Advanced Virgo and the two Advanced LIGO detectors.
The signals are injected in 1 s long segments with a sampling rate of $8192$~Hz. Since the signals are shorter than 1 s, they are padded with zeros to fill the length. We use \texttt{Bilby}'s dynesty sampler with $2000$~live points. 

To measure the quality of our inferred posteriors we compute the Bayes factor, $\mathcal{B}$,
defined as
\begin{eqnarray}
  \log \mathcal{B} &=& \log P(d|H) - \log P(d|H_0) 
  \nonumber 
  \\
  &=& -\frac{2}{T} \sum_{\rm IFOs}\sum_{i=1}^{N/2} \left (
  \frac{|\hat s_i - \hat h_i|^2 - |\hat s_i|^2}{S_{n\,i}} \right ) \,,
\end{eqnarray}
where $P(d|H)$ is the evidence of the signal and $P(d|H_0)$ is the evidence of the noise, defined as the expectation of the observed data given the hypothesis that there is no signal (null hypothesis, $H_0$). A value of $\log \mathcal{B}<0$ implies that the observed data is more consistent with being noise than a signal; however, $\log \mathcal{B}>0$, does not immediately imply that there is a matching signal, but rather just that the model is a better one than Gaussian noise. 
In a network with several detectors, the terms corresponding to the different detectors are added so that if all of them are positive/negative the value of $\log \mathcal{B}$ gets more positive/negative. Therefore, adding more detectors to the network increases the confidence in the validation of the hypothesis.
In the next section we discuss the significance of the values of $\log \mathcal{B}$ in terms of the matched-filter signal to noise ratio (SNR) of the detectors network, which is defined as \cite{jaranowski2012gravitational}
\begin{equation}
{\rm network\,\, SNR} = \sqrt{\sum_{\rm IFOs} {\rm SNR}^2}.
\end{equation}
SNR here is the optimal
SNR of each detector defined as
\begin{equation}
  {\rm SNR} = \sqrt{\frac{4}{T}\sum_{i}^{N/2} \frac{|\hat h_i|^2}{S_{ni}}},
\end{equation}
with $\Delta t$ being the sampling time of the signal.

\subsection{Error estimation}
\label{sec:error}

Assuming supernovae behave in nature as they do in simulations, a detected waveform will differ from a template waveform for two reasons. First, our template is an imperfect model of the simulation results. Second, the measured signal will contain contributions from detector noise. Here we propose a method of quantifying both.

We differentiate between the intrinsic parameters of the master template, $\Theta_{\rm int, MT} = \{f_{\rm peak, MT}, (\Dhs)_{\rm MT} \}$
and the real values of the waveform, $\Theta_{\rm int, real} = \{f_{\rm peak, real}, (\Dhs)_{\rm real}\}$ (i.e., the signal actually measured in the detector). There will be error in the inferred MT parameters due to both the imperfect template and detector noise. The posterior probability for the inference of the real parameters is
\begin{equation}
  p(\Theta_{\rm real}| d, H) = p(\Theta_{\rm int, real}|\Theta_{\rm int, MT}, H') 
   p(\Theta_{\rm MT}| d, H)
\end{equation}
where $\Theta_{\rm real}$ and $\Theta_{\rm MT}$ are the full sets of parameters including the corresponding
intrinsic parameters. $p(\Theta_{\rm int, real}|\Theta_{\rm int, MT}, H')$ is the probability of a waveform having intrinsic parameters $\Theta_{\rm int, real}$ given a match to a template with parameters
$\Theta_{\rm int, MT}$ and a model $H'$ for the relation between $\Theta_{\rm int, real}$ and $\Theta_{\rm int, MT}$ (naturally, we use as model ($H'$) that $\Theta_{\rm int, real} = \Theta_{\rm int, MT}$.). This is precisely the term including the information about the error associated with the master template itself.

{$p(\Theta_{\rm int, real}|\Theta_{\rm int, MT}, H')$ can be estimated using the Bayes' theorem 
as a function of the likelihood $p(\Theta_{\rm int, MT}|\Theta_{\rm int, real})$. Given that we have no a priori information about the prior probability $p(\Theta_{\rm int, MT})$ we use a uniform distribution  and hence $p(\Theta_{\rm int, real}|\Theta_{\rm int, MT}, H') \propto p(\Theta_{\rm int, MT}|\Theta_{\rm int, real}, H')$. To compute the likelihood of the MT parameters given the real,  $p(\Theta_{\rm int, MT}|\Theta_{\rm int, real}, H')$, we follow the procedure in Appendix~\ref{app:errors}.
We perform this analysis using only the $402$ waveforms of~\cite{richers2017equation} and not including the $12$ signals from Table~\ref{tab:CCSNModels}. 
The result is that this model is consistent with the data and that the likelihood (and hence the posterior distribution) can be modelled as a normal distribution with a mean given by the model $H'$ and with standard deviations $\sigma_{f}=0.024\cdot f_{\rm peak}$ and $\sigma_\Delta=0.065\cdot \Dhs$ for $f_{\rm peak}$ and $\Dhs$, respectively.

In practice, this means that we can perform our inference algorithm using the master templates as described in Section~\ref{sec:bayes} and then add the contribution of the additional error to the posterior distribution. In order to propagate error, we generate a random number $N_{\rm spawn}$ of additional values for each sample of the original posterior distribution using the values of the original sample as mean and the standard deviation values given above. The collection of values form a blurred  posterior distribution including an estimate of the errors of the master template. We have tried values of $N_{\rm spawn}=1,10,50$ but all produce very similar posterior distributions. Therefore, in 
this work we use $N_{\rm spawn}=1$ to not increase the computational cost of the analysis.

 One alternative to this procedure would be to estimate directly the physical parameters of the system (and their errors) using the relations suggested by \cite{richers2017equation} between $\rho_{\rm c}$ and $f_{\rm peak}$, and between $T/|W|$ and $\Dh$. In this case we want to infer a set of physical parameters $\Theta_{\rm int, phys}$, so the procedure is identical as the one described above but using $\Theta_{\rm int, phys}$ instead of $\Theta_{\rm int, real}$. The detailed analysis can be found in Appendix~\ref{app:errors}. 
In this case we use as a model $H'$ the result of the linear fits of the parameters $\rho_c$ and $T/|W|\sin^2 \theta$ as a function of the master template parameters $f_{\rm peak}$ and $\Dhs$, respectively, which results in:
\begin{align}
  &\rho_{\rm c} = \left (7.3\times\frac{f_{\rm peak}}{\rm 1000 Hz} - 1.67 \right) \times 10^{14} {\rm g\,cm}^{-3}, \label{eq:fit1}\\
  &T/|W|\sin^2 \theta = (1.1\times \Dhs + 17 ) \times 10^{-4}. \label{eq:fit2}
\end{align}
Since the amplitude of the waveform depends on the inclination angle $\theta$, 
we can put constraints only on the combination $T/|W| \sin^2 \theta$, but not on $T/|W|$.
The likelihood, $p(\Theta_{\rm int, phys}|\Theta_{\rm int, MT})$ can be modelled as a normal distribution with mean given by the values above and standard deviation $\sigma_{\rho_{\rm c}}=0.07\cdot \rho_{\rm c}$ and $\sigma_{T/W}=0.08\cdot T/|W|$.

Another alternative would be to directly use the estimated error for the master template (see Sect.~\ref{sec:MasterTemplate}) to add the uncertainty at the time of sampling, i.e. our model would return a waveform randomly drawn from the distribution obtained for the master template. That would incorporate automatically these systematic errors in the posteriors of the Bayesian inference. This interesting possibility may be explored in our future work.

\begin{figure}[h!]
 \includegraphics[width=0.48\textwidth]{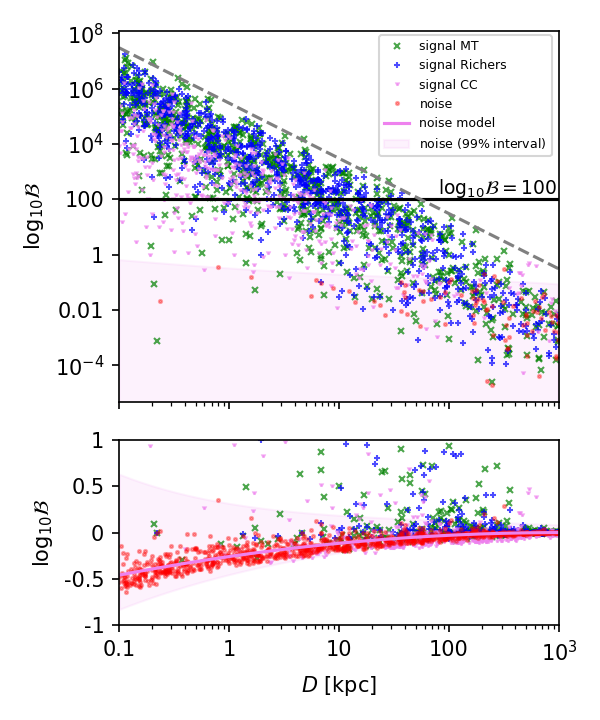}
 \caption{Logarithm of the Bayes factor as a function of the distance of the injected signal. Both signal (blue and green markers) and noise (red dots) injections are plotted. The violet solid line (only visible in the bottom panel since it has negative values) and violet area display the mean value for noise injections and the $99\%$ confidence interval (see main text for details on the noise model). The grey dashed line in the top plot shows the $1/D^2$ dependence. The lower panel shows a detailed view of the region with low values of the Bayes factor in linear scale.} 
 \label{fig:Bayes}
\end{figure}

\begin{figure}[h!]
 \includegraphics[width=0.48\textwidth]{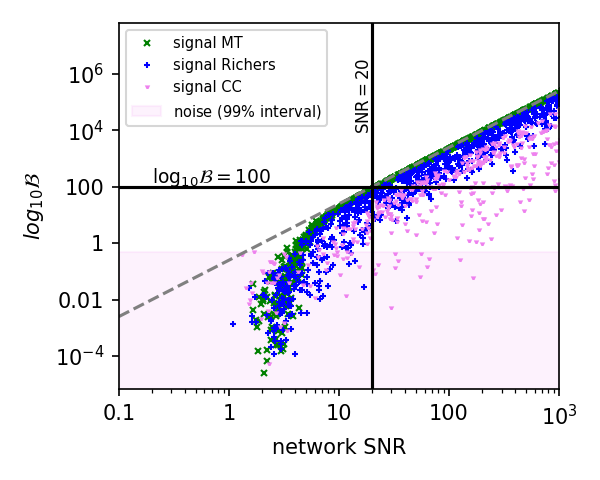}
 \caption{Logarithm of the Bayes factor as a function of the network SNR for different types of injections. The dashed grey line shows a quadratic dependence with the network SNR. The violet area corresponds to values of $\log_{10} \mathcal{B}<0.5$, which approximately corresponds to the $99\%$ confidence interval of the noise injections. }
 \label{fig:BayesSNR}
\end{figure}

\subsection{Waveform injection}

In order to test systematically our parameter estimation method we perform a series of injections under different conditions with random values of the parameters. Each of the series consists of $1000$ injections in randomly generated Gaussian noise of a 3-detector network corresponding to the two Advanced LIGO detectors and Advanced Virgo at design sensitivity \cite{SensitivityCurve}. Injections are performed at random locations in the sky (constant probability per solid angle in the sky), random luminosity distance, $D$, in the range $0.1-1000$~kpc, random inclination angle, $\theta$, of the rotation axis (constant probability per solid angle) and random polarization angle, $\psi$. Signals from 3D simulations (see Section~\ref{sec:CCSNWaveforms}) depend on an additional azimuthal angle with respect to the rotation axis, $\phi$, describing the orientation of the source. In those cases we a use random angle in the range $[0,2\pi]$. This angle can not be inferred from our analysis as the templates do not contain this dependence. While $\phi$ only produces a significant effect on the waveforms after bounce since the bounce dynamics is mainly axisymmetric, we vary $\phi$ for good measure. For each of the injections we perform the Bayesian inference analysis described in Sections~\ref{sec:bayes} and \ref{sec:error} for the parameters $f_{\rm peak}$, $\Dhs$ and $\psi$ assuming that the location in the sky, distance and time of bounce are known.

We carry out four main series of injections depending on the type of waveforms selected: 

\begin{itemize}
  \item {\it Noise}: Null injections with zero amplitude that serve as a reference for cases in which there is only noise and no signal.
  \item {\it Signal MT}: Waveforms using the master template to generate signals with random values of $f_{\rm peak}$ in the range $[600,1000]$~Hz and $D \cdot \Delta h$ in the range $[0,700]$~cm.
  \item {\it Signal Richers}: Waveforms chosen ramdomly among the 402 waveforms selected from the catalog of~\cite{richers2017equation} using the procedure of Section~\ref{sec:MT}.
  \item {\it Signal CC:} Waveforms chosen randomly among the $12$ core collapse signals of Table~\ref{tab:CCSNModels} as discussed in Section~\ref{sec:CCSNWaveforms}.
\end{itemize}


\section{Results}
\label{sec:results}

\begin{figure}[t]
 \centering
 \includegraphics[width=0.48\textwidth]{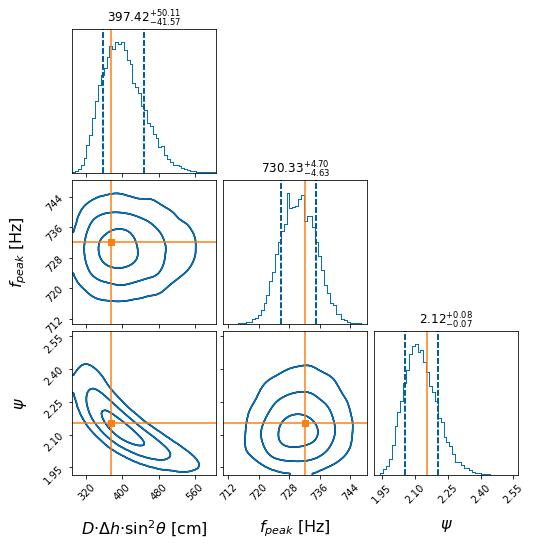}
 \caption{Corner plot of the posterior probability distribution 
 for a master template injection. The network SNR of the signal is $27.5$ and $\log_{10} \mathcal{B}=157$. Contours show $1$-$\sigma$, $2$-$\sigma$ and $3$-$\sigma$ levels. Dashed blue lines show the $1$-$\sigma$ confidence intervals, and orange lines correspond to the injected values. The correlation between $\psi$ and $D\Delta h\cdot\sin^2\theta$ is a result of the particular sky location and detector orientation.}
 \label{fig:CornerMT}
\end{figure}

\begin{figure*}[t!]
 \includegraphics[height=0.36\textwidth]{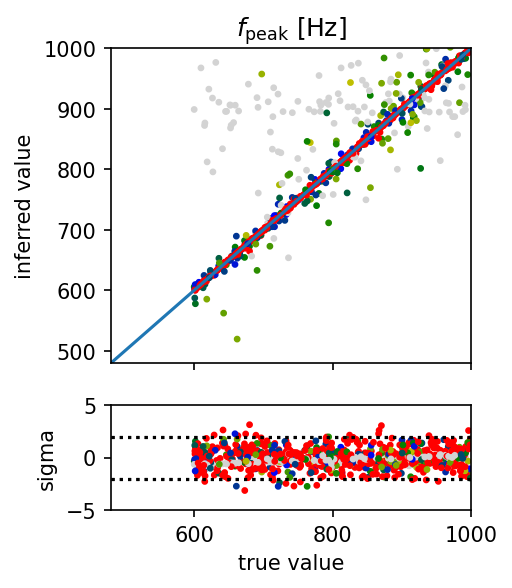}
 \includegraphics[height=0.36\textwidth]{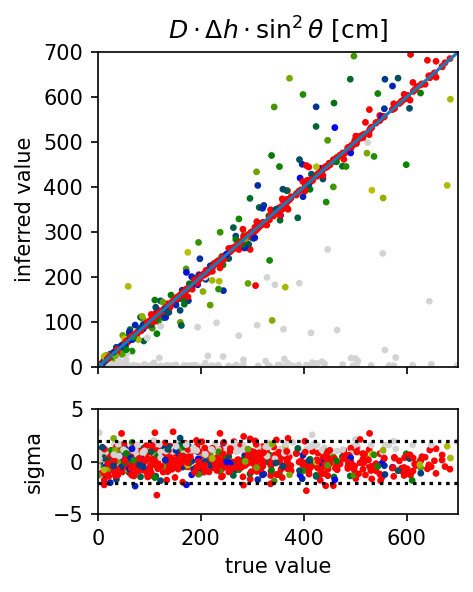}
 \includegraphics[height=0.36\textwidth]{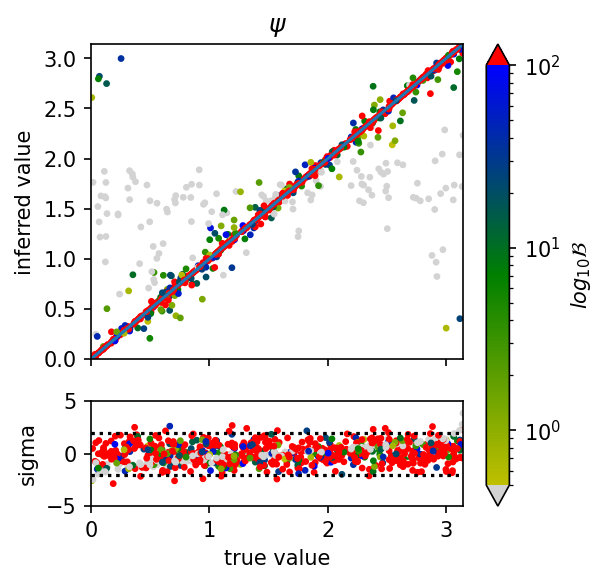}
 \caption{Bayesian inference for master template injections. {\it Upper panels:} Median of the inferred posteriors for $f_{\rm peak}$ (left panel), $\Dhs$ (middle panel) and $\psi$ (right panel) as a function of the true value of the injected waveform. Colors indicate the logarithm of the Bayes factor. Values of $\log_{10} \mathcal{B}>100$ are indicated in red and $\log_{10} \mathcal{B}<0.5$ in grey. The blue solid line in the diagonal of all plots corresponds to equal true and inferred values. {\it Lower panels:} error in the inferred values in term of standard deviations. Horizontal dashed lines indicate the $2$-$\sigma$ interval.  }
 \label{fig:MT}
\end{figure*}

\begin{figure}[ht]
 \includegraphics[width=0.48\textwidth]{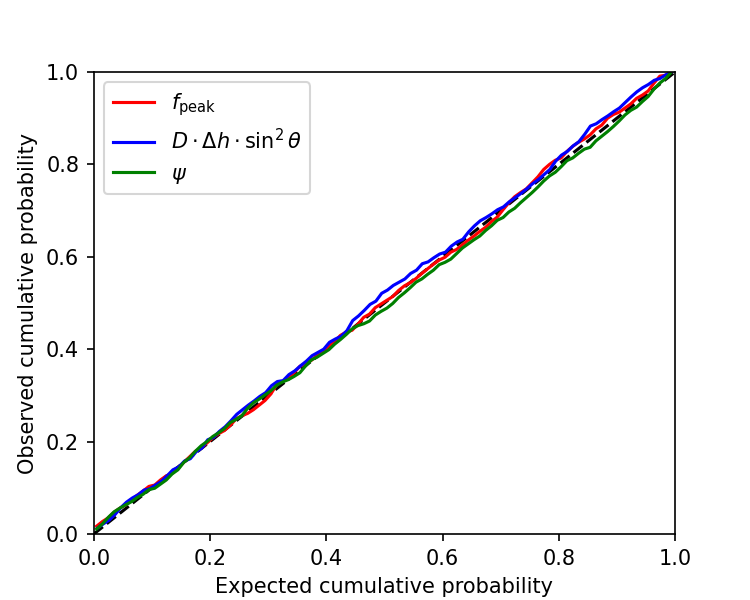}
 \caption{P-P plot for master template injections: Observed cumulative probability distribution of the deviation of the inferred values from the injected values vs the expected cumulative distribution when comparing the posterior distribution of each injection with the corresponding injected value. For this analysis we consider a set on injections in which the parameter distribution of the injections match the priors. Dashed black line indicates equal probability.}
 \label{fig:PP_MT}
\end{figure}

\subsection{Significance of the inferred values}

Before discussing in detail each of the different injection cases separately we first consider in more general grounds the quality of the inferred values by computing the Bayes factor, $\mathcal{B}$. Fig.~\ref{fig:Bayes} shows the dependence of $\log_{10} \mathcal{B}$ for the different series of injections as a function of the distance to the source. For signal injections, the maximum of $\log_{10} \mathcal{B}$ approximately scales with $1/D^2$ (dashed grey line). This upper limit corresponds to injections with optimal inclination and sky location, as well as the highest amplitude $D\cdot\Delta h$. Below this line we find injections with non-optimal configurations and lower amplitudes. Note that even for small distances some signals can have very low values of the Bayes factor. Since the injected values of $f_{\rm peak}$ are close to the maximum sensitivity of the detectors, the dependence of the sensitivity on frequency is fairly flat and does not produce any observable systematic influence in the value of $\log_{10} \mathcal{B}$, for the range of frequencies considered.

To try to interpret the value of the Bayes factor, and in particular to understand which signals would be detectable and which would be hidden in the detector noise, we perform a series of injections with pure noise.
For those injections we obtain values of $\log_{10} \mathcal{B}$ close to zero. There is a small dependence on the distance to the source, which appears because the distance (which is considered to be known in the analysis) affects the priors. For large distances, the expected amplitude of the signal is small, therefore the analysis cannot conclude whether there is an unobserved signal in the noise or if it is instead pure noise, which results in a $\log_{10} \mathcal{B}\approx 0$. For small distances the model predicts a high signal amplitude for most orientations. Therefore, the latter case, there is a tendency to obtain $\log_{10} \mathcal{B} < 0 $, i.e.~to reject the presence of a signal given that there is only noise. To model the effect of the detector noise we fit $\log_{10} \mathcal{B}$ as a function of $\log D$ with a second-order polynomial (violet line in Fig.~\ref{fig:Bayes}). Next we perform a linear fit of the quadratic difference between this noise model and the values for different injections; this linear fit gives us an estimate of the variance of $\log_{10} \mathcal{B}$ as a function of $\log D$. Since the distribution is clearly non Gaussian, instead of using this variance directly, we rescale it by a constant value to estimate the 0.5th and 99.5th percentiles. The resulting $99\%$ confidence interval (i.e.~the region interval within which a signal is consistent with noise) is displayed as a violet area in Fig.~\ref{fig:Bayes}. In those cases the posteriors obtained from the Bayesian analysis are uninformative and essentially follow the priors, except for $\Dhs$, which tends to be close to zero. 
\begin{table}[ht]
 \centering
 \caption{Fraction of injections of the different series with a logarithm of the Bayes factor above a certain threshold. Threshold SNR values are estimated for the MT signals, though Richers and CC injections have SNR values below this threshold.}
 \begin{tabular}{c@{\hskip 15pt}c@{\hskip 15pt}c@{\hskip 15pt}c}
  \hline\hline
   Injection & $\log_{10} \mathcal{B}>0.5$ &$\log_{10} \mathcal{B}>16$& $\log_{10} \mathcal{B}>100$\\ 
   series& (SNR$>2$) & (SNR$>8$)& (SNR$>20$) \\\hline
   MT & $0.64$ & $0.53$ & $0.44$ \\
   Richers & $0.67$ & $0.57$ & $0.48$ \\
   CC & $0.51$ & $0.41$  & $0.32$ \\
   noise & $0$ & $0$ & $0$ \\ \hline\hline
 \end{tabular}
 \label{tab:Bayes}
\end{table}
In general, any signal with $\log_{10} \mathcal{B}>0.5$ will fulfill the criterion of being above the noise. However, this does not imply that those signals are detectable or that one could extract meaningful information. We also note that this threshold could increase significantly if we were to use real detector noise, in which non-stationary noise (glitches) is present. Although a proper analysis of the detectability of this kind of signals is out of the scope of this paper, we can make an estimate based on the SNR. Previous studies on the performance of the cWB pipeline \cite{Abbott:2020,Szczepanczyk2021}
for CCSN waveforms (in particular for waveforms similar to those considered here) have shown that typically more than $50\%$ of the events with SNR$>20$ are detectable when real detector noise is considered~\citep{Szczepanczyk2021}. However, in this work we place the threshold on $\log_{10}\mathcal{B}$ since it offers a more clear separation between noise and true signal in our tests. In Fig.~\ref{fig:BayesSNR} we show the dependence of the Bayes factor on the network SNR. The maximum value of $\log_{10} \mathcal{B}$ is proportional to the square of the network SNR (shown as a dashed grey line). All injections resulting in $\log_{10} \mathcal{B} >100$ have a SNR$>20$. Therefore, for practical purposes, we will use this threshold on the value of the Bayes factor as a conservative estimate of the detectability of a signal based only on observed data (i.e., assuming high-SNR, low-$\mathcal{B}$ signals to be undetectable). 

If it were possible to detect CCSN signals using (optimal) matched-filtering techniques, the detection threshold would lower to SNR$>8$, which  is satisfied for all signals with $\log_{10} \mathcal{B} >16$. Table~\ref{tab:Bayes} shows the fraction of injections that pass the various thresholds considered for the different series of injections performed. There can be interpreted as an approximate comparison of the detectability of the different classes of waveforms. However, the fraction of detectable waveforms depends on the distribution of inherent amplitudes of the generated waveforms, which is not uniform across the datasets. The Richers dataset has a slightly higher fraction of detectable events than the MT dataset because it skews more to high-amplitude events, even though the injected MT signals are generated from the same template as is used in the inference procedure. However, the mismatch with the CC waveforms is large enough to significantly decrease the number of detectable events.

\subsection{Master template signal injections}
\label{sec:results:MT}

In the case of master template injections, the templates used for the inference and for the injections are obviously the same. This means that, for the appropriate parameters, the template is a perfect match for the waveform injected. As a result, the accuracy of the inferred values will improve as the SNR of the injected signal increases. Figure~\ref{fig:CornerMT} shows an illustrative example of the posterior probability distribution obtained for a case with a relatively high SNR of $27.5$. We are able to recover the parameters of the injected signal within the $1$-$\sigma$ confidence interval with a relatively low error ($0.3\%$, $11\%$ and $1.4\%$ for $f_{\rm peak}$, $\Dhs$ and $\Psi$, respectively). 

Figure~\ref{fig:MT} shows the median values and errors of the inferred posteriors depending on the injected values for the three quantities of the analysis. For signals with high significance ($\log_{10} \mathcal{B}>100$, red dots) the relative difference between the inferred median and injected values is almost always smaller than $2.3\%$, $14\%$ and $7\%$ for $f_{\rm peak}$, $\Dhs$ and $\psi$, respectively. Only in one case the inferred value of $\Dhs$ has a larger relative error ($\sim 40\%$, see the outlier red dot in the middle panel of Fig.~\ref{fig:MT}) albeit within the 3-$\sigma$ limit, which is expected for $1000$ injections. The error in the median values are mostly confined within the $2$-$\sigma$ confidence interval.

To evaluate the quality of the posterior distributions we show in Fig.~\ref{fig:PP_MT} a P-P plot.
For this plot we performed an ad hoc set of $1000$ injections with the same distribution of the  parameters $\Dhs$, $f_{\rm peak}$ and $\psi$ as the priors used for the Bayesian analysis.
This plot compares the observed cumulative probability distribution of the deviation of the inferred values from the injected ones (vertical axis) with the expected cumulative distribution obtained when comparing the posterior distribution of each injection with the corresponding injected value. A straight line (equal probability) indicates that the posterior describes properly the results obtained, i.e.~it is a good estimator of the error associated with the inference process. Deviations from this equal probability line would indicate an excess or scarcity of events at certain confidence level. The results obtained for master template signals show that the posterior follows closely the equal probability line for all quantities and therefore the analysis can be used with high confidence in its results.

\subsection{Richers et al. signal injections}
\label{sec:results:Richers}

\begin{figure}[t]
 \includegraphics[width=0.48\textwidth]{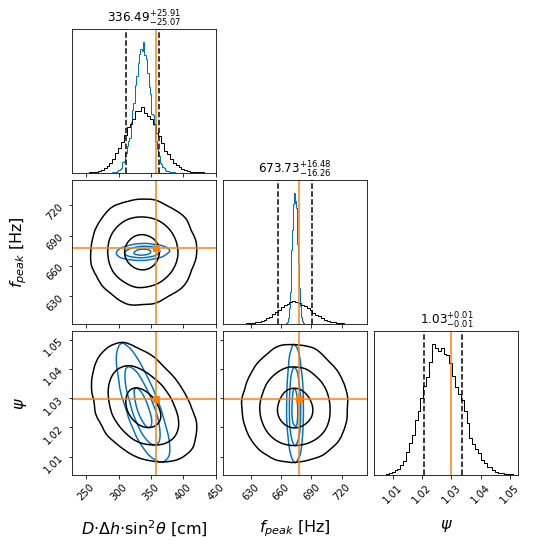}
 \caption{Corner plot of the posterior probability distribution 
 for an example injection of the Richers et al. catalog. The network SNR of the signal is $57$ and $\log_{10} \mathcal{B}=463$. 
 Blue contours show the posterior distribution computed by the Bayesian analysis and black contours, the corresponding distribution after incorporating the error in the template. Contours show $1$-$\sigma$, $2$-$\sigma$ and $3$-$\sigma$ levels. Dashed black lines display the $1$-$\sigma$ confidence intervals and orange lines correspond to the injected values.}
 \label{fig:cornerRic}
\end{figure}

\begin{figure*}[t!]
 \includegraphics[height=0.36\textwidth]{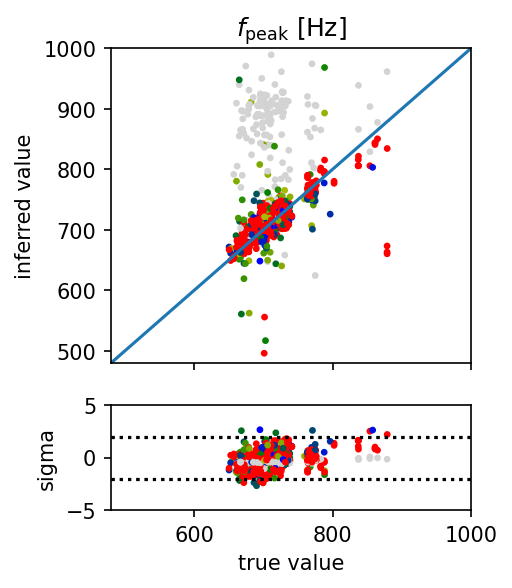}
 \includegraphics[height=0.36\textwidth]{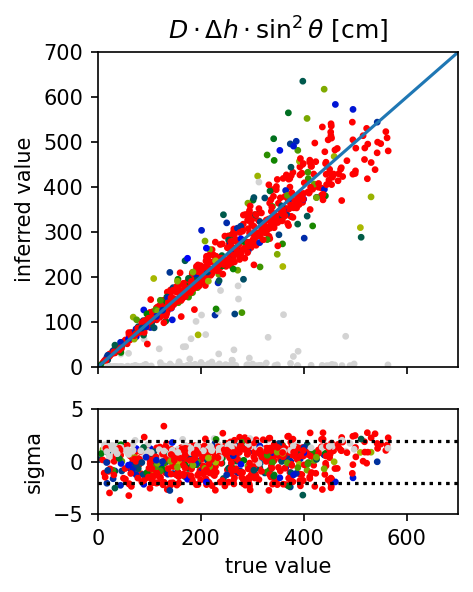}
 \includegraphics[height=0.36\textwidth]{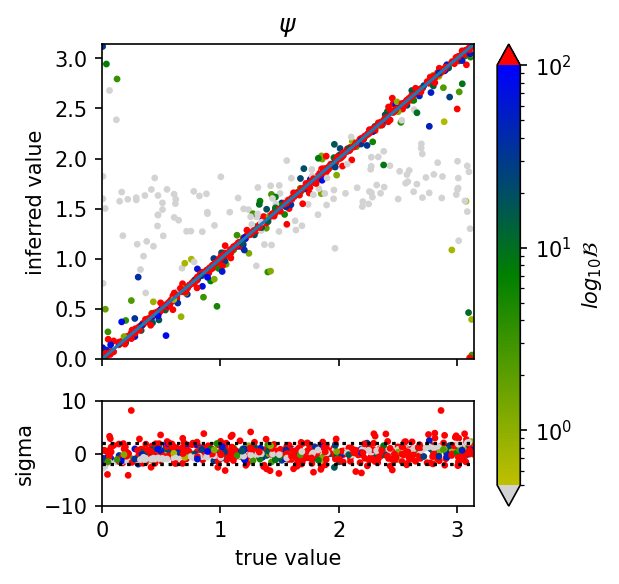}
 \caption{Bayesian inference of $f_{\rm peak}$ , $\Dhs$  and $\psi$  as a function of the true value of the injected waveform, when using injections from the Richers et al. catalog. 
 See Fig.~\ref{fig:MT} for details.}
 \label{fig:Ric}
\end{figure*}

\begin{figure*}[ht]
\includegraphics[width=0.425\textwidth]{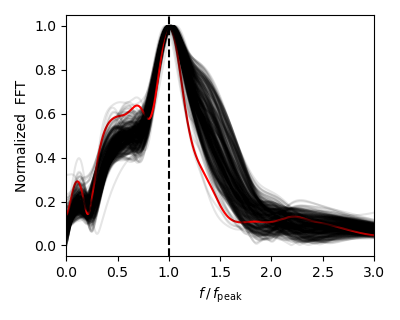}
 \includegraphics[width=0.425\textwidth]{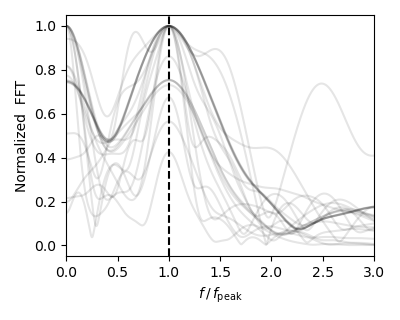}
 \caption{Normalized Fourier transform of the bounce signal (between $5$~ms before to $6$~ms after bounce) as a function of the frequency normalized to the peak frequency, for all the waveforms used from the Richers et al catalogue (right) and for all the waveforms from Table~\ref{tab:CCSNModels} (left). The latter show in general a richer structure with multiple peaks compared to the former. Highlighted in red the waveform \texttt{A634w3.00\_SFHo\_ecapture\_0.1}, mentioned in the text, which also has a secondary peak. }
 \label{fig:doublepeak}
\end{figure*}

Next we discuss our results when injecting waveforms from the Richers et al. catalog~\cite{richers2017equation}. In this case we have to post-process the posteriors given by the Bayesian inference algorithm to take into account the intrinsic error of the master templates (see Section~\ref{sec:error}). Fig.~\ref{fig:cornerRic} shows an illustrative example of the posterior probability of a loud signal with a SNR of $57$. The plot shows both the posterior probability obtained directly with the Bayesian inference process (blue) and after post-processing the posteriors to incorporate the error of the master template (black). Since the SNR is high, the predicted error associated with the detector noise (blue) is relatively low and the master template error has a significant contribution to the final posteriors of $f_{\rm peak}$ and $\Dhs$ (black). Note that $\psi$ is unaffected. For this particular case, the injected value falls at the edge of the $3$-$\sigma$ confidence interval before the inclusion of the template errors but within $1$-$\sigma$ after its inclusion. At sufficiently high SNR the contribution of noise diminishes and the final posteriors are essentially dominated by the master template errors (the transition seems to happen at $\mathrm{SNR}\sim 20$ in this work). Correspondingly, at sufficiently low SNR the noise becomes dominant to the point that the posterior is insensitive to the template error contribution. 

Fig.~\ref{fig:Ric} depicts the inferred values depending on the injected values and their errors. The observed behaviour for all three parameters is similar to the case of master template injections discussed before. However, we note now a larger dispersion of the inferred values due to the intrinsic error of the templates. Of the five outliers visible in $f_{\rm peak}$ (top panel, red circles) three of them correspond to injections of the same waveform (\texttt{A634w3.00\_SFHo\_ecapture\_0.1}, according to the naming convention in \cite{richers2017equation}), which is a model with artificially decreased electron capture rates. The Fourier transform of this waveform presents a secondary peak at about the frequency that is recovered by our analysis (see Fig.~\ref{fig:doublepeak}). That hints that, for some models, secondary modes may be excited at bounce that would need to be included in the model for more accurate results.

\begin{figure*}[t!]
 \includegraphics[height=0.355\textwidth]{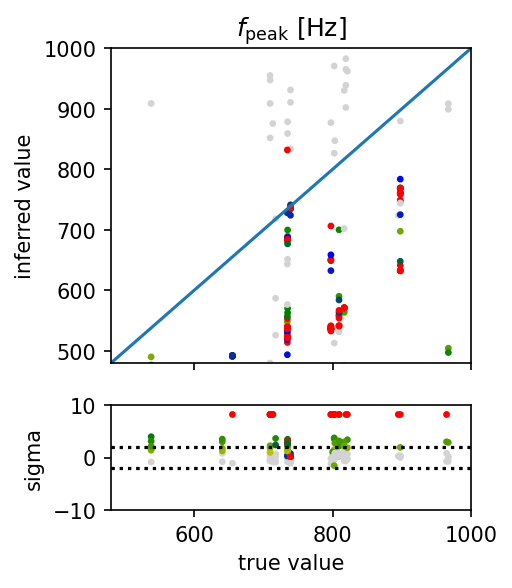}
 \includegraphics[height=0.355\textwidth]{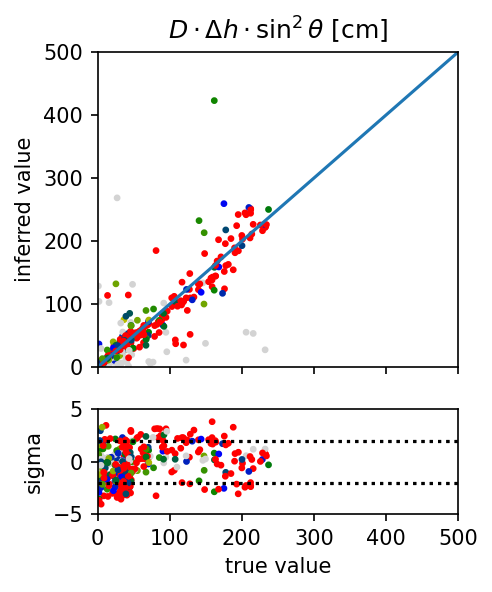}
 \includegraphics[height=0.355\textwidth]{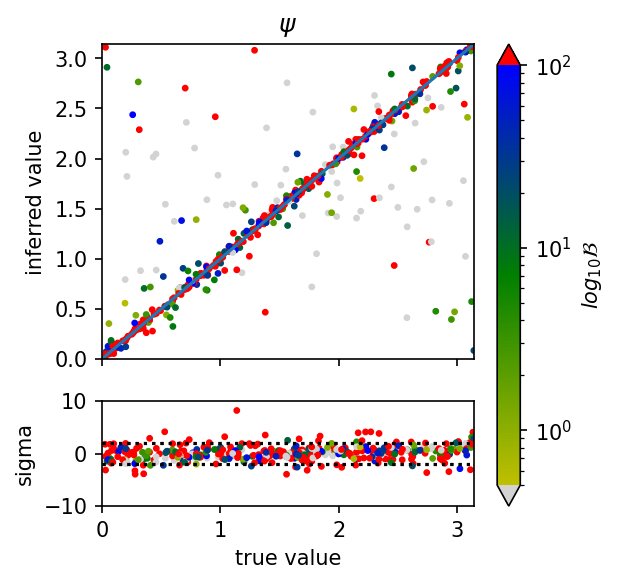}
 \caption{Bayesian inference of $f_{\rm peak}$, $\Dhs$ and $\psi$ as a function of the true value of the injected waveform, when using injections from Table~\ref{tab:CCSNModels}. The description of the plots is as in Fig.~\ref{fig:MT}}
 \label{fig:NR}
\end{figure*}

The errors shown in Fig.~\ref{fig:Ric} lay mostly in the $2$-$\sigma$ interval, although the number of injections outside this interval is larger than for the master template case. This could be an effect of the assumption that the error due to the master template follows a normal distribution (see Section~\ref{sec:error}). This could be corrected by using a more accurate description for that distribution and should be the subject of future studies.

\subsection{Core collapse signal injections}

\begin{figure*}[ht]
\includegraphics[width=0.425\textwidth]{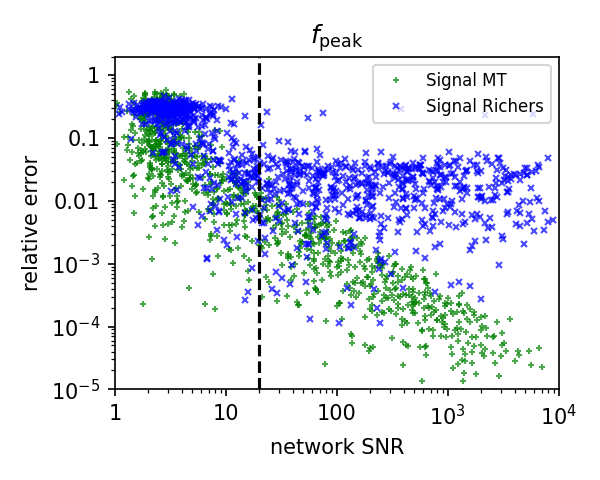}
 \includegraphics[width=0.425\textwidth]{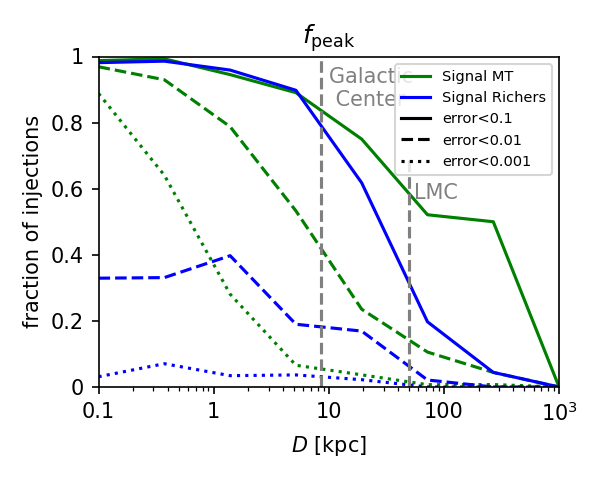}\\
 \includegraphics[width=0.425\textwidth]{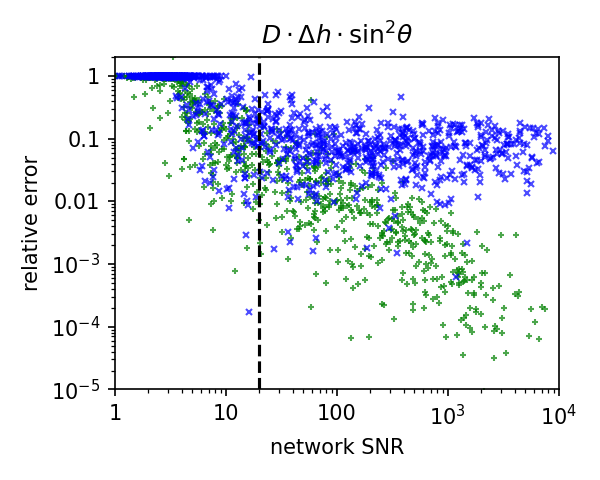}
 \includegraphics[width=0.425\textwidth]{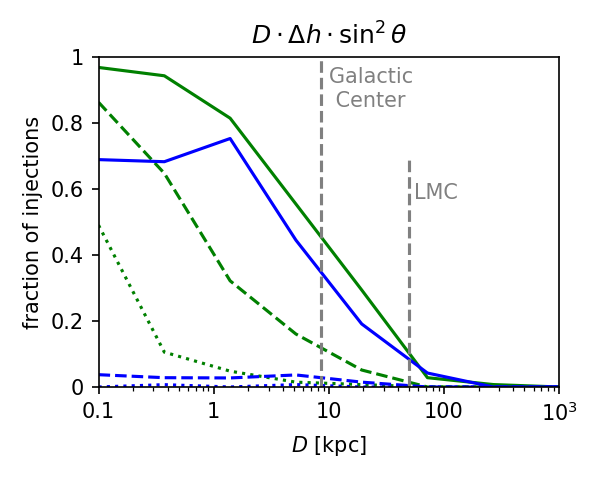}\\
 \includegraphics[width=0.425\textwidth]{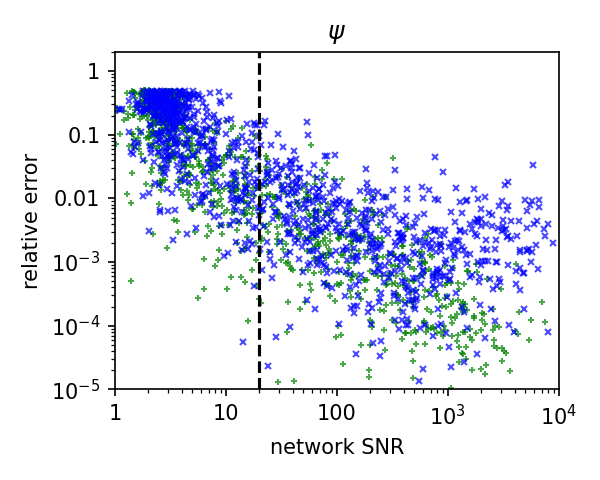}
 \includegraphics[width=0.425\textwidth]{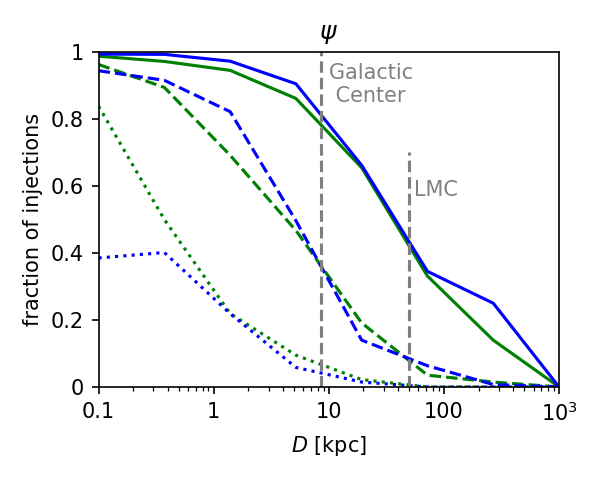}\\
 \caption{Left panels: Relative error in the inference of $f_{\rm peak}$ (top), 
 $\Dhs$ (middle), and $\psi$ (bottom) as a function of the network SNR for both, master template injections (green symbols) and injections from the Richers et al. catalog~\cite{richers2017equation} (blue symbols). Dashed vertical lines mark the approximate detection threshold of SNR$=20$.  Right panels: Fraction of injections at a given distance with a relative error smaller 
 than a certain threshold ($10\%$, $1\%$ and $0.1\%$) for the three parameters studied.}
 \label{fig:fractionError}
\end{figure*}

We turn now to consider signal injections with the waveforms from Table~\ref{tab:CCSNModels}, i.e.~waveforms which were not employed to build the master template. The inferred values and errors are displayed in Fig.~\ref{fig:NR}. In this case, there are clear mismatches between the inferred and the true values.
The worst results are obtained for 
$f_{\rm peak}$, for which the inferred value is systematically underestimated. This is an extreme case of the pathology commented in the previous section affecting to most signals instead of only a few: The post-bounce evolution of the CCSN signals from Table~\ref{tab:CCSNModels} is more complex than those from~\cite{richers2017equation} and cannot be modelled so faithfully 
by our master template waveform (see Fig.~\ref{fig:doublepeak}). In particular, the Fourier transform of the strain near bounce presents multiple peaks, whereas the Richers waveforms generally only show one. The presence of  additional peaks have been discussed in \cite{Fuller2015}, and may be related to overtones of the fundamental $f$-mode, to non-linear couplings or to the presence of $g$-modes. 
This produces a double confusion in the algorithm: i) it is difficult to evaluate what the nominal injected value of $f_{\rm peak}$ is; and ii) the Bayesian algorithm could be fitting the wrong oscillation mode. This is a strong indication that further work is needed to converge on a realistic bounce signal and/or our master template needs improvement possibly by adding secondary frequencies. 

Regarding $\Dhs$, the results seem to follow the right behaviour for values of $\Dhs<250$. Above that point, however, the inferred value is systematically underestimated. The results for $\psi$ are better but in many cases there are still large deviations. Both effects could be attributed to the strong mismatch in $f_{\rm peak}$, but the presence of additional unaccounted modeling dependencies on these two parameters cannot be dismissed. This highlights the potential for future investigations to systematically examine these dependencies, while also considering possible improvements to the waveform models.

\subsection{Dependence with SNR and distance}

We next discuss the dependence of the results on the network SNR and on the distance to the source. In so doing we only consider the master template and the Richers et al. waveforms, leaving out of the analysis the additional CCSN waveforms of Table~\ref{tab:CCSNModels} because of the poor match between these waveforms and the template. Figure~\ref{fig:fractionError} shows both the relative error of the inferred values as a function of the network SNR (left panels) and the fractions of injections below a certain relative threshold as a function of the distance to the source (right panels). As a reference we show in the right panels the distance to the Galactic Center ($\sim 8.5$~kpc) and to the Large Magellanic Cloud ($49.6$~kpc, \cite{Pietrzynski2019}) that hosted SN 1987A.

For master template injections the relative error (green symbols in left panels) decreases with increasing SNR, as expected. The green lines in the right panels show the fraction of injections at a given distance that are recovered within a given error (regardless of the Bayes factor). For a Galactic event ($D\sim 10$~kpc), both $f_{\rm peak}$ and $\Dhs$ could be recovered with less than $10\%$ error for $\sim 80\%$ and $\sim 60\%$ of the cases, respectively. For the case of $f_{\rm peak}$ this accuracy could be maintained for half of the events up to $\sim 100$~kpc. For signals with low significance (low value of $\log_{10} \mathcal{B}$), the analysis recovers systematically very low values of $\Dhs$. This is because the observation is compatible with noise, i.e.~with an observation of a zero-amplitude
waveform. 

For injections from the Richers et al. catalog, the relative error (blue symbols in left panels) decreases with the network SNR in the regime in which it is dominated
by the detector noise (network SNR$<20$, vertical dashed line). However, for values of the network SNR associated with an unambiguous detection (network SNR$>20$) the relative error remains fairly constant and it is dominated by the error in the template. This limits the accuracy of the analysis to a few percent for $f_{\rm peak}$ and to $\sim 10\%$
for $\Dhs$. For a Galactic event, it is still possible to infer $f_{\rm peak}$ with a relative error smaller than $10\%$ (see the green curves in the top right panel of Fig.~\ref{fig:fractionError}). 
However, the accuracy of the recovery of $\Dhs$ degrades, 
with about $25\%$ of the events showing a relative error larger than $10\%$. Given that $\psi$ is mostly unaffected by the master template error, the relative error keeps decaying at larger SNR, saturating at values larger than $\sim 100$, consistent with the discussion in Section~\ref{sec:results:Richers}.

Regarding the dominant source of error in the inference process, we compare the injections in which only the detector noise is considered in the analysis with those which also incorporate the systematic error in the template. The metric that we use for this comparison is the ratio of the sizes of the $1$-$\sigma$ uncertainty intervals with and without including template errors, minus one. Values larger (smaller) than one indicate that the template (noise detector) error dominates. At large distances (low SNR) the error is dominated by detector noise, while at short distances (high SNR) it is dominated by the template systematic error. The transition point can be determined as the distance or SNR at which the error in half of the injections is dominated by detector noise. For measurements of $f_{\rm peak}$ ($\Dhs$), we find that template systematic errors are dominant within $11$~kpc ($4$~kpc) or for network SNR higher than $22$ ($68$).
At a typical galactic distance of $10$~kpc, the systematic error dominates in $60\%$ and $17\%$ of the cases for $f_{\rm peak}$ and $\Dhs$, respectively. Therefore, systematic errors in the templates are very relevant for a galactic event and should not be neglected.

\begin{figure}[ht]
 \includegraphics[width=0.48\textwidth]{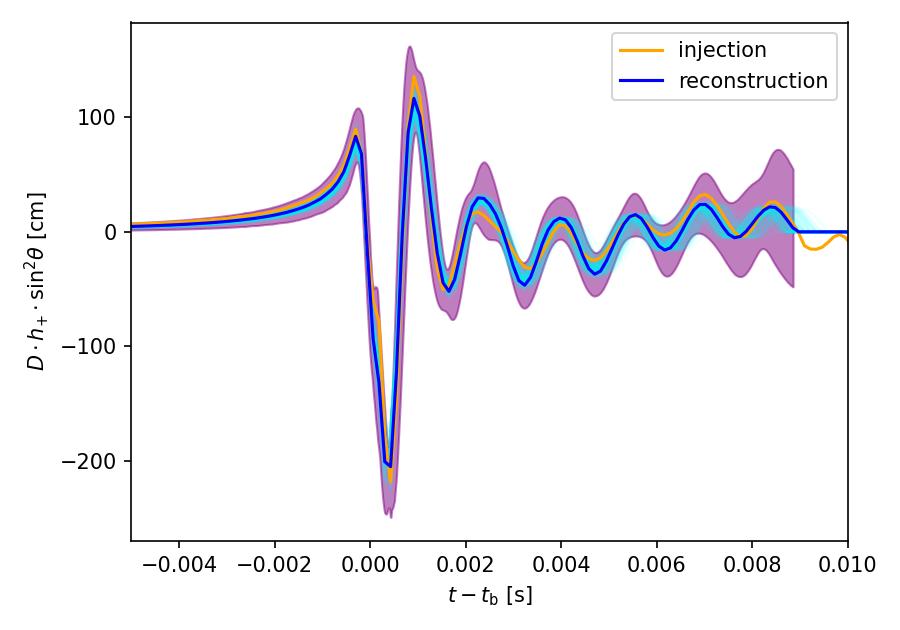}
 \caption{Reconstructed (blue) and injected (orange) signal for the same case shown in Fig.~\ref{fig:cornerRic}. Purple region is the $2$-$\sigma$ confidence interval estimated with the error of the master templates. Cyan curves correspond to waveforms generated with $100$ random samples of the posterior distribution (including template errors).}
 \label{fig:WFMreconstruction}
\end{figure}

\subsection{Waveform reconstruction}

\begin{figure*}[t!]
 \includegraphics[height=0.4\textwidth]{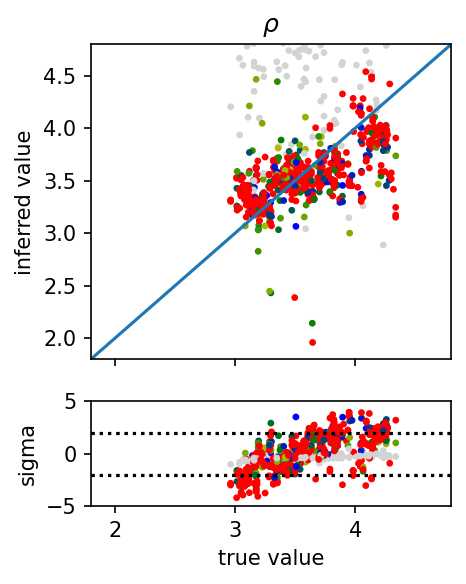}
 \includegraphics[height=0.4\textwidth]{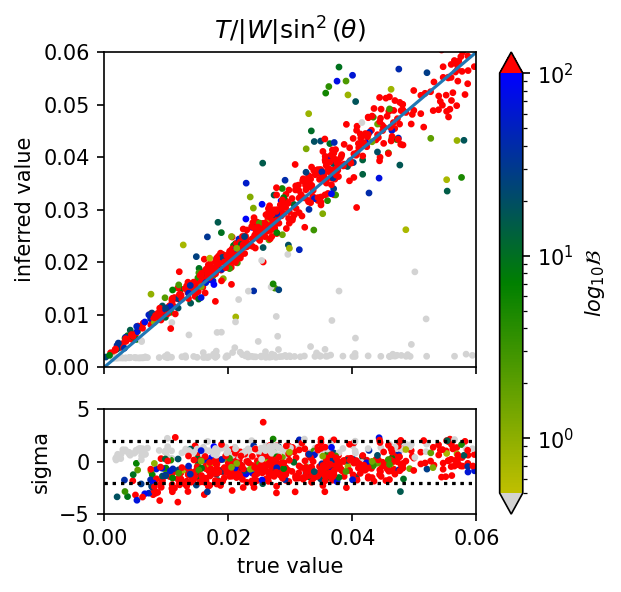}
 \caption{Bayesian inference of $\rho_{\rm c}$ and $T/|W|\sin^2\theta$  as a function of the true value of the injected waveform, when using injections from the Richers et al. catalog~\cite{richers2017equation}. See Fig.~\ref{fig:MT} for details.}
 \label{fig:NR2}
\end{figure*}

The determination of the parameters of a signal allows reconstruction of the injected signal using the median values of the posterior distribution to generate a waveform using the master template. Fig.~\ref{fig:WFMreconstruction} shows an example of waveform reconstruction from the Richers et al. catalog (same example as in Fig~\ref{fig:cornerRic}). We have estimated the error of the waveform in two ways: First, using the standard deviation of the master template (see Section~\ref{sec:MT}) we have evaluated the $2$-$\sigma$ confidence interval (purple region). This estimation only considers the master template error and works well for this particular case with a high network SNR of $57$. In general, however, it underestimates the error for cases dominated by detector noise errors (lower SNR). Second, we plot a selection of master template waveforms (cyan curves) generated using $100$ random samples from the posterior distribution. This procedure could be used to evaluate confidence intervals in a more accurate way, incorporating both template and detector noise errors. In both cases the injected template is within the error interval.

\subsection{Inference of physical parameters}

Finally, we use the procedure outlined in Section~\ref{sec:error} to estimate the physical parameters $\rho_{\rm c}$ and $T/|W|\sin^2\theta$ from the waveform. This estimation is exclusively carried out for the waveforms in the Richers et al. catalog. The master template waveforms do not come from numerical simulations and so they do not have associated physical parameters. There is only partial information about the CC waveforms publicly available, so while we could still perform the analysis, we would not be able to compare with the injected values. 
Furthermore, we need bounce signals consistent with the template in order to reliably infer the physical parameters $\rho_{\rm c}$ and $T/|W|\sin^2\theta$, which is not the case for these models as shown in ~Section~\ref{sec:CCSNWaveforms}.

Fig.~\ref{fig:NR2} shows the inferred values versus the injected ones. Regarding $\rho_{\rm c}$, the inferred values have a large scatter around the true values. The combined effect of the errors coming from the detector noise, the master template errors and the dispersion in the $\rho_{\rm c}$ vs $f_{\rm peak}$ relation, results in a large scatter of the inferred values in the range $3-4\times 10^{14}$~g/cm$^3$ over the entire range of the injected ones. 
Moreover, the fact that the range of possible values for $\rho_{\rm c}$ is relatively narrow does not help, raising the degree of accuracy necessary to obtain a meaningful physical result. 

The results are significantly better for $T/|W|\sin^2\theta$. The dispersion is larger than in the case of $\Dhs$ due to the imperfect mapping between waveform amplitude and rotation rate, but its value can be recovered with $\sim 25\%$ accuracy in most cases. Notice that the value that we infer includes the dependence on $\theta$, which is not a physical parameter of the system. If one desires to remove the dependence on $\theta$ without any previous knowledge of its value, then the results obtained become automatically lower limits. Note that a low Bayes factor event (grey dots) is equivalent to a non-detection. In that case, the inferred values are usually zero, meaning that the only information that we can extract from a non-detection is that $T/|W|\ge 0$, i.e.~it is completely uninformative. 
In Section~\ref{sec:conclusions} we discuss possible ways of determining $\theta$ independently to break the existing degeneracy.

\section{Conclusions and outlook}
\label{sec:conclusions}

In this work we have proposed a procedure to infer proto-neutron star properties from future gravitational-wave observations. We have focused on the early-time (bounce) GW signal from the collapse of stellar cores for fast rotating progenitors. Despite the complex and stochastic character of CCSN signals, the bounce part of the signal is rather regular and depends on a simple form on two parameters: the bounce amplitude, $\Dh$, and the peak frequency, $f_{\rm peak}$. The main interest of these two quantities is that their values correlate with the physical properties of the PNS and, in particular, with the ratio of rotational-kinetic energy to potential energy, $T/|W|$, and the central density, $\rho_{\rm c}$, at bounce~\cite{richers2017equation}. The main goal of this investigation has been to provide estimates for such two parameters directly from the bounce GW signal.

We have developed a simple parametric waveform template (master template) to model the bounce. This template has been constructed using a carefully selected set of 402 models from the Richers et al. waveform catalog~\cite{richers2017equation}, which globally comprises over $1800$ 
axisymmetric simulations extending up to about $10$~ms of post-bounce evolution.
Using the master template, we have performed Bayesian inference on signals injected in Gaussian colored noise of a three-detector network of advanced GW interferometers (Advanced LIGO and Advanced Virgo). We have performed this analysis making use of the Bayesian inference library \texttt{Bilby}~\cite{Ashton_2019}. We have been able to propagate the errors due to the master template inaccuracies into the final posterior distributions, generating error estimates for the inferred values that include Gaussian detector noise and model uncertainties.

Our procedure has been tested with four data sets: i) null injections (pure noise), ii) master template injections, iii) bounce signals from \cite{richers2017equation} and iv)  other CCSN waveforms from the literature. Injected template waveforms are recovered with better accuracy than injected waveforms from the training set, which in turn are recovered better than injected waveforms from other simulations. While master injections are only affected by detector noise error, more realistic signals are affected by the fact that the master template itself is an imperfect representation of simulation waveforms. Using the bounce signals of \cite{richers2017equation}, we have been able to recover the injected parameters, namely $f_{\rm peak}$, $\Dhs$, and $\psi$; with an accuracy better than $10\%$ for more than $50\%$ of the detectable events. While for waveforms outside of the training catalog the accuracy and confidence of the results worsens, it is still possible to obtain valid information for sufficiently loud signals.
Inferences of the physical parameters of the PNS, $T/|W|$ and $\rho_{\rm c}$, have similarly low accuracies for the same reason. Since modeling uncertainty plays such a large role in the otherwise very regular signals, there is a significant opportunity for future work to address these modeling uncertainties with relatively short-duration simulations. 

We have identified the main limitations of our procedure that presently are preventing us  from reaching a higher accuracy for inferred parameters. Addressing the following issues will be subject of future investigations:
\begin{itemize}
 \item We model the errors due to the master template inaccuracies with a simple normal distribution. This may be responsible for an excess of injections outside the $2$-$\sigma$ confidence interval. The solution is relatively simple and would consist of using the complete distribution that we obtain from the analysis of the errors. 
 \item In some of the the CC bounce waveforms (i.e.~not in the catalog of \cite{richers2017equation}) we have identified multiple oscillating frequencies in the bounce signal that cannot be modeled by a single peak frequency. It is unclear whether the origin of these other peaks is modeling deficiencies or differing choices of initial conditions. Therefore, it may be necessary to incorporate additional degrees of freedom in the templates (e.g.~a secondary frequency) and converge on unified waveforms produced by multiple codes. A more detailed analysis of the bounce signal under more realistic conditions (e.g.~treatment of deleptonization, heavy nuclei, progenitor profiles, fully dynamical spacetime) and different numerical methods would be necessary to implement this improvement.
 \item All the waveforms in the Richers et al. catalog employed to create the master template make use of the same $12 M_\odot$ progenitor star. Using waveforms from a wider variety of progenitors (e.g.~\cite{Mitra:2023}) could allow for the creation of master templates that better approximate realistic CCSN waveforms from any core collapse with significant rotation. Additionally, we would obtain improved error estimates and better estimators of the physical properties.
\end{itemize}

One of the main caveats of our analysis is the degeneracy of the measurement of $\Dh$ (or $T/|W|$) with the value of the inclination angle $\theta$. In order to break this degeneracy we would need an independent measurement of constraint on the value of $\theta$. Some possibilities are:
\begin{itemize}
 \item It has been shown \cite{Hayama2016} that the stokes parameters of the gravitational wave signal of the post-bounce evolution carry information about the PNS rotation if non-axisymetric deformations are present. These deformations could appear systematically in the form of bar mode or spiral instabilities for sufficiently high rotation rates \cite{Ott2005,Ott2007,Scheidegger2010,Takiwaki2016,Shibagaki2020,Bugli2023,Pan2021}. In this case it should be possible to constrain the inclination angle by measuring the ratio of the different stokes parameters.
 \item Long term observations of the supernova remnant could allow to measure the velocity field of the ejecta and their asymmetries. If those asymmetries are significant and related to the presence of strong rotation (e.g.~bipolar flows), the inclination angle could be estimated.
 \item Very quickly rotating progenitors are a possible source of long-GRBs. In those cases the collimated jet would producing the GRB is expected to be aligned with the angular momentum of the system. Therefore its observation could constrain the inclination angle. 
\end{itemize}

While our results provide a possible way to infer PNS properties from GW observations of CCSNe, they should be taken with care as there are a number of simplifying assumptions in our model that could have an impact on the inferred waveforms and physical parameters should they be relaxed. Perhaps the most important simplification has been the use of zero-mean, coloured Gaussian noise into which the injections have been made. An immediate extension of the present work would be to account for actual (non-Gaussian) detector noise. In addition, we should explore the impact in the analysis on the number of detectors in the network (e.g.~with only the two aLIGO detectors it may not be possible to recover the polarization angle) or the use of next generation detectors such as the Einstein Telescope\cite{Punturo:2010zza}, Cosmic Explorer\cite{reitze2019cosmic} or NEMO\cite{Ackley2020}. Results incorporating those elements will be presented elsewhere.

\section{Acknowledgements}
We thank Christopher Berry, Sylvia Biscoveanu, Marie-Anne Bizouard, Jade Powell, and Marek Szczepanczyk for their useful comments and suggestions. This research has been supported by the Spanish Agencia Estatal de Investigaci\'on (grant PID2021-125485NB-C21 funded by MCIN/AEI/10.13039/501100011033 and ERDF A way of making Europe). Further support is provided by the Generalitat Valenciana (Prometeo program for excellent research groups grant CIPROM/2022/49 and Astrophysics and High Energy Physics program grant ASFAE/2022/003 funded by MCIN and the European Union NextGenerationEU (PRTR-C17.I1)), by the EU's Horizon 2020 research and innovation (RISE) programme H2020-MSCA-RISE-2017 (FunFiCO-777740), and by the European Horizon Europe staff exchange (SE) programme HORIZON-MSCA-2021-SE-01 (NewFunFiCO-101086251). 
PCD acknowledges support from the {\it Ramon y Cajal} funding (RYC-2015-19074). SR was supported by a NSF Astronomy \& Astrophysics Postdoctoral Fellowship under
Grant No. 2001760. EA was supported by RK MES grant No. AP13067834 and NU Faculty Development Grant No. 11022021FD2912. We thank the YITP for hospitality and support during the 2019 long-term workshop
{\it Multi-Messenger Astrophysics in the Gravitational Wave Era} during which the idea for this project was discussed.


\appendix
\section{Computation of the errors associated to the master template}
\label{app:errors}

\begin{figure*}[ht]
 \includegraphics[width=0.48\textwidth]{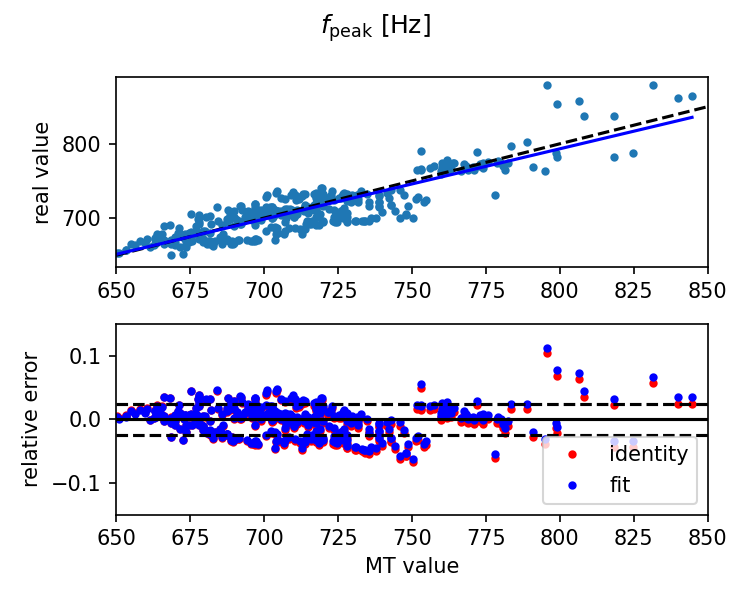}
 \includegraphics[width=0.48\textwidth]{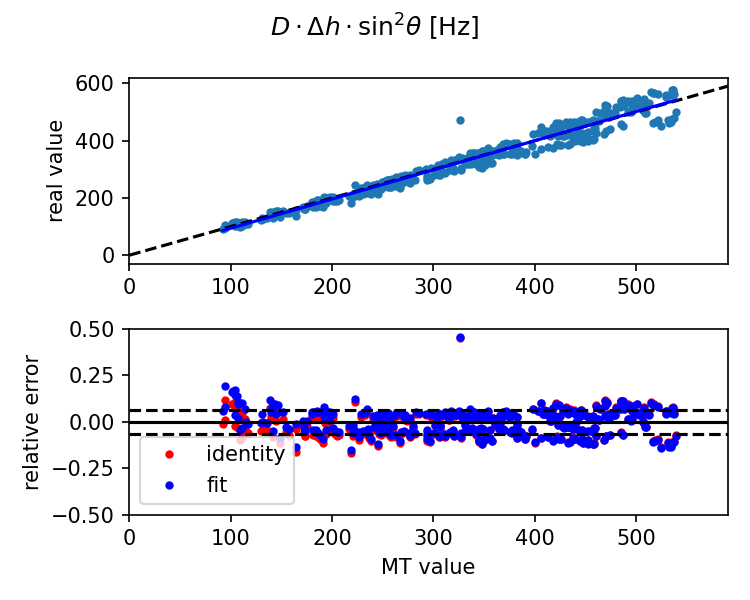}
 \caption{Upper panels show the parameters $\Theta_{\rm int, real}$ for each template (real value) as a function of the value of the maximum likelihood of $\Theta_{\rm int, MT}$ (MT value) for this template, for the parameters $f_{\rm peak}$ (left panel) and $\Dhs$ (right panel). Dashed lines indicate equal value of the MT and real values (identity model) and the blue line, the result of a linear regression (fit model). Lower panels show the corresponding relative errors considering the the identity and fit models. Dashed lines show $2$-$\sigma$ confidence interval for the fit model.}
 \label{fig:errors}
\end{figure*}
\begin{figure*}[ht!]
 \includegraphics[width=0.48\textwidth]{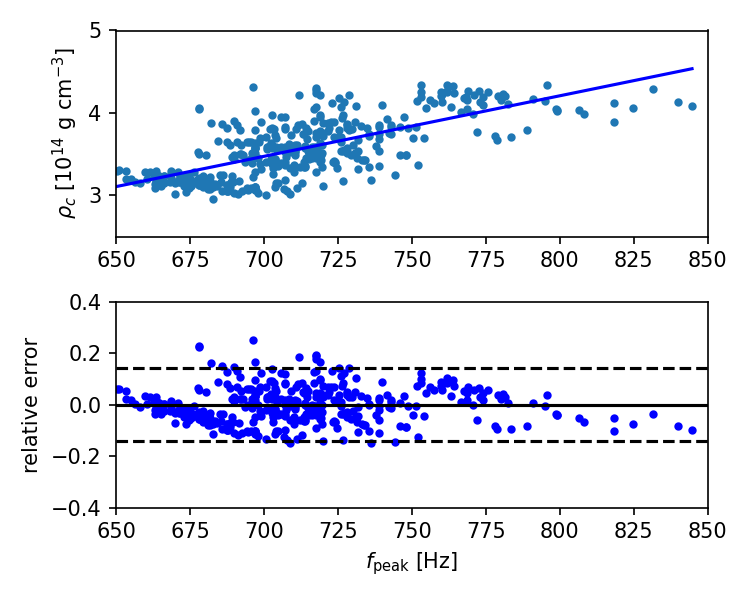}
 \includegraphics[width=0.48\textwidth]{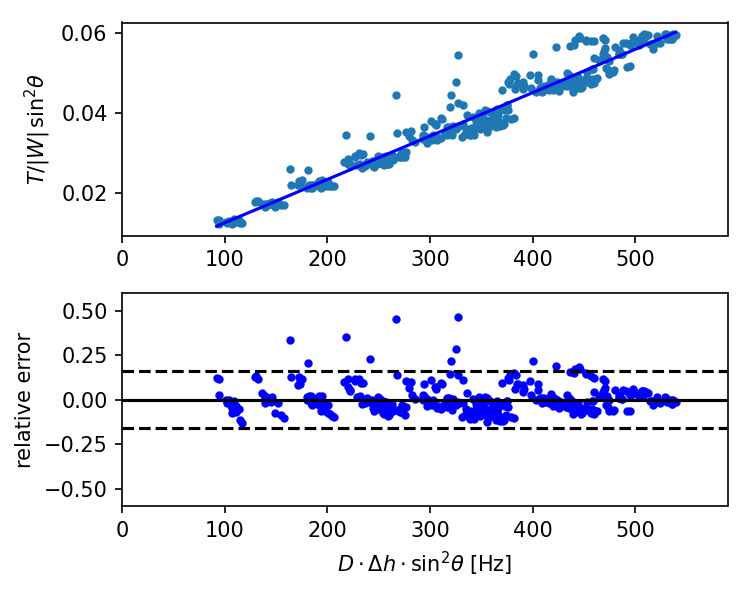}
 \caption{Upper panels show the values of the maximum likelihood of $\Theta_{\rm int, MT}$ (MT value) for a template with parameters $\Theta_{\rm int, phys}$ (physical value), for the parameters $\rho_c$ (left panel) and $T/|W|\sin^2\theta$ (right panel). The blue line shows the result of a linear regression. Lower panels show the corresponding relative errors with respect to the fit. Dashed lines show the $2$-$\sigma$ confidence interval for the fit. The visible groups of points correspond to models with the same initial rotation rate but different rotation law. }
 \label{fig:errors2}
\end{figure*}

As described in Section~\ref{sec:results:MT}, we need to relate the values inferred
for the master template, $\Theta_{\rm int, MT}$, with the real values, $\Theta_{\rm int, real}$,
by means of the probability $p(\Theta_{\rm int, MT}|\Theta_{\rm int, real}, H')$. 
In order to compute this probability, first we find the master template parameters, $\Theta_{\rm int, MT}$, that match best with each waveform from the Richers et al. catalog, i.e. those parameters that maximize the waveform likelihood given by
\begin{equation}
 \log \mathcal{L}_{\rm waveform} = -\frac{\sum_{i=1}^{N/2} |\hat h^{\rm MT}_i - \hat h^{\rm Ric}_i|^2}{\sum_{i=1}^{N/2} |\hat h^{\rm Ric}_i|^2}.
  \label{eq:error_likelihood1}
\end{equation}
Second, we consider that the resulting maximized parameters for all the waveforms in the catalogue constitute samples of the distribution that we want to measure, $p(\Theta_{\rm int, MT}|\Theta_{\rm int, real}, H')$.

Note that the denominator of Eq.~\eqref{eq:error_likelihood1} is just a normalization constant and does not affect the computation of the maximum. The estimation has been performed using $\theta=0$, $\psi=0$, $D=10$~kpc; however, the analysis does not depend on the arbitrary choice of these extrinsic parameters. We compute the maximum likelihood using the Nelder-Mead simplex algorithm \cite{Gao2012} implemented in the \texttt{SciPy} python library \footnote{https://scipy.org}.

Fig.~\ref{fig:errors} shows the relation between the value of $\Theta_{\rm int, MT}$ computed as the maximum likelihood for each waveform and the real values in the template. We have performed linear regressions for 
both parameters $f_{\rm peak}$ and $\Dhs$, and in both cases they result in good fits ($R$ values of 
$0.88$ and $0.98$, respectively) with a slope compatible with unity. We test two alternatives as a model for $H'$: one case in which we assume that the slope is one (identity model, with $\Theta_{\rm int, real} = \Theta_{\rm int, MT}$) and a second model in which we use the result of the fit (fit model). The relative errors with respect to the models can be found in the lower panels of Fig.~\ref{fig:errors}. In both cases errors are very similar, so we will consider the identity model hereafter for simplicity. The distribution of relative errors for both variables can be modelled as a normal distribution with zero mean and standard deviations $0.024$ and $0.065$ for $f_{\rm peak}$ and $\Dhs$, respectively. Dashed lines in lower panels show the $2$-$\sigma$ confidence interval. 

Note that for the case of $\Dhs$, there is an outlier at $4.7$ sigmas from the mean. The corresponding waveform has a high frequency artifact close to the maximum amplitude, which could be related to a numerical artifact. In any case, removing this outlier has very little impact on the results of the analysis so we have kept it for simplicity.

To compute the errors associated to the estimation of the physical parameters $\Theta_{\rm int, phys}$ from the template parameters $\Theta_{\rm int, MT}$, we assume, following \cite{richers_sherwood_2016_201145}, a linear dependence of $\rho_{\rm c}$ and $T/|W|\sin^2 \theta$ on $f_{\rm peak}$ and $\Dhs$, respectively. We compute the parameters of the linear function by performing fits to the data, as shown in Fig.~\ref{fig:errors2}. The results of the fits, which serve as model $H'$, are:
\begin{align}
 &\rho_{\rm c} = \left (7.3\times\frac{f_{\rm peak}}{\rm 1000 Hz} - 1.67 \right) \times 10^{14} {\rm g\,cm}^{-3}, \label{eq:fit1_appendix}\\
  &T/|W|\sin^2 \theta = (1.1\times \Dhs + 17 ) \times 10^{-4},
  \label{eq:fit2_appendix}
\end{align}
with $R$ values $0.71$ and $0.97$, respectively. 
The relative error with respect to this model (lower panels in Fig.~\ref{fig:errors2}) can be modelled as normal distributions with standard deviations $0.07$ and $0.08$ for $\rho_{\rm c}$ and $T/|W|\sin^2\theta$, respectively.

A closer inspection of the model for $T/|W|\sin^2 \theta$ reveals that there are $4$ outliers outside the $4$-$\sigma$ confidence interval. The numerical simulations corresponding to these four outliers were performed using an artificially increased electron capture rate (a factor 10) that was used by \cite{richers2017equation} to test the influence of this parameter on the waveform. \cite{richers2017equation} showed that the electron capture rate treatment can change the slope of the relation between $T/|W|$ and the peak amplitude (see their Fig. 16), therefore modifying the slope of Eq.~\eqref{eq:fit2_appendix}. However, they showed that the effect of the EOS in the slope is much smaller (see their Fig. 6). In this sense, it is reasonable to think that using models with much tighter constrained electron capture rates would lead to a better model for $T/|W| \sin^2 \theta$ with no outliers present. 

Regarding the model for $\rho_c$, the low value of $R$ is related to the large scatter in the data. This is a clear indication that the relation in this case may be depending on other parameters. In particular, \cite{richers2017equation} pointed out the influence of the EOS on the slope of Eq.~\eqref{eq:fit1_appendix} (see their Fig. 8, bottom panel).

For the current work we will consider the data as it is without further restrictions. This will lead to more conservative results and larger errors than in the case of a particular EOS and a model for electron capture rates.

\bibliography{biblio}

\end{document}